\documentclass[12pt,letterpaper,onecolumn,amsfonts,amssymb,amsmath,floatfix,
floats,nobalancelastpage]{revtex4}

\usepackage{graphicx}
\begin{document}

\title{
Desynchronization and Speedup
in an Asynchronous Conservative Parallel Update Protocol
}
\author{A. Kolakowska}
\email{alicjak@bellsouth.net}
\author{M. A. Novotny}
\email{man40@ra.msstate.edu}
\affiliation{ \baselineskip=0.42truecm
Department of Physics and Astronomy, and ERC
Center for Computational Sciences,
P.O. Box 5167, Mississippi State, MS 39762-5167}

\date{\today}

\begin{abstract}
\centerline{\bf ABSTRACT}
\baselineskip=0.42truecm
In a state-update protocol for a system of $L$ asynchronous 
parallel processes that communicate only with nearest neighbors, 
global desynchronization in operation times can be deduced from kinetic 
roughening of the corresponding virtual-time horizon (VTH). The utilization 
of the parallel processing environment can be deduced by analyzing the microscopic 
structure of the VTH. In this chapter we give an overview of how the methods 
of non-equilibrium surface growth (physics of complex systems) can be applied to 
uncover some properties of state update algorithms used in distributed parallel 
discrete-event simulations (PDES). In particular, we focus on the asynchronous 
conservative PDES algorithm in a ring communication topology. The time evolution 
of its VTH is simulated numerically as asynchronous cellular automaton whose update 
rule corresponds to the update rule followed by this algorithm. There are two cases 
of a balanced load considered: (1) the case of the minimal load per processor, 
which is expected to produce the lowest utilization (the so-called worst-case 
performance scenario); and, (2) the case of a general finite load per processor. 
In both cases, we give theoretical estimates of the performance as a function of $L$ 
and the load per processor, i.e., approximate formulas for the utilization 
(thus, the mean speedup) and for the desynchronization (thus, the mean memory 
request per processor). It is established that the memory request per processor, 
required for state savings, does not grow without limit for a finite number 
of processors and a finite load per processor but varies as the conservative 
PDES evolve. For a given simulation size, there is a theoretical upper bound for 
the desynchronization and a theoretical non-zero lower bound for the utilization. 
We show that the conservative PDES are generally scalable in the ring communication 
topology. The new approach to performance studies, outlined in this chapter, 
is particularly useful in the search for the design of a new-generation of algorithms that 
would efficiently carry out an autonomous or tunable synchronization.
\end{abstract}

\maketitle


\section{Introduction \label{intro}}

Parallel discrete-event simulations (PDES) are a technical tool to uncover 
the dynamics of information-driven complex systems. Their wide 
range of applications in contemporary sciences and technology \cite{Fuj00} 
has made them an active area of research in recent years. Parallel and 
distributed simulation systems constitute a complex system of their own, 
whose properties can be uncovered with the well-established tools of 
statistical physics.

In PDES physical processes are mapped to logical processes (assigned to 
processors). Each logical process manages the state of the assigned 
physical subsystem and progresses in its {\it local virtual time} (LVT). 
The main challenge arises because logical processes are not synchronized 
by a global clock. Consequently, to preserve causality in PDES the algorithms 
should incorporate the so-called local causality constraint \cite{CM79,Fuj90} 
whereby each logical process processes the received messages from 
other processes in non-decreasing time-stamp order. Depending on the way the 
local causality constraint is implemented, there are two broadly defined classes 
of update protocols \cite{Fuj00}: conservative algorithms 
\cite{CM79,Mis86,Lub87,Lub88} and optimistic algorithms 
\cite{Jef85,DR90,PS92,Ste93,FC94}. In conservative PDES, an algorithm does 
not allow a logical process to advance its LVT until it is certain that 
no causality violation can occur. In the conservative update protocol a logical 
process may have to wait to ensure that no message with a lower time stamp 
is received later. In optimistic PDES, an algorithm allows a logical process 
to advance its LVT regardless of the possibility of a causality error. 
The optimistic update protocol detects causality errors and provides a 
recovery procedure from the violation of the local causality constraint 
by rolling back the events that have been processed prematurely.

There are several aspects of PDES algorithms that should be considered in 
systematic efficiency studies. Some important aspects are: the synchronization 
procedures, the utilization of the parallel environment as measured by the 
fraction of working processors, memory requirements which may be assessed by 
measuring the statistical spread in LVTs (i.e., the desynchronization), 
inter-processor communication handling, scalability as measured by evaluating 
the performance when the number of computing processors becomes large, 
and the speedup as measured by comparing the performance with sequential 
simulations. In routinely performed studies to date, the efficiency is 
investigated in a heuristic fashion by testing the performance of a selected 
application in a chosen PDES environment, i.e., in a parallel simulator. 

Only recently a new approach in performance studies has been introduced 
\cite{KTNR00,KNTR01,KNRGT01,KNKG02,KNK03,KNR03,KNGTR03,TKNG03,GKTN04} 
in which the properties of the algorithm are examined in an abstract way, 
without a reference to a particular application platform. In this approach 
the main concept is the simulated {\it virtual time horizon} (VTH) defined 
as the collection of LVTs of all logical processes. The evolution rule of 
this VTH is defined by the communication topology among processors and by 
the way in which the algorithm handles the advances in LVTs. The key assumption 
here is that the properties of the algorithm are encoded in its representative 
VTH in analogy with the way in which the properties of a complex system are 
encoded in some representative non-equilibrium interface. In this way, fundamental  
properties of the algorithm can be deduced by analyzing its 
corresponding simulated VTH.

In this chapter we give an overview of how the methods of non-equilibrium surface 
growth (physics of complex systems) can be applied to uncover some properties of 
state update algorithms used in PDES. In particular, we focus on the asynchronous 
conservative PDES algorithm in a ring communication topology, where each processor 
communicates only with its immediate neighbors. The time evolution of its VTH is 
simulated numerically as an asynchronous cellular automaton whose update rule 
corresponds to the update rule followed by this algorithm. The purpose of this study 
is to uncover generic properties of this class of algorithms.

In modeling of the conservative update mode in PDES, we represent sequential events on 
processors in terms of their corresponding LVTs. A system of processors or 
{\it processing elements} (PE) is represented as a one-dimensional grid. 
The column height that rises above the $k$-th grid point is a 
building block of the simulated VTH and represents the total time of 
operations performed by the $k$-th processor. These operations can be 
seen as a sequence of update cycles, where each cycle has two phases. 
The first phase is the processing of 
the assigned set of discrete events (e.g., spin flipping on the assigned 
sublattice for dynamic Monte Carlo simulation of lattice systems). 
This phase is followed by a messaging phase that closes the cycle, when a processor 
broadcasts its findings to other processors. But the messages broadcasted by other 
processors may arrive any time during the cycle. Processing related to these messages 
(e.g., memory allocations/deallocations, sorting and/or other related operations) 
are handled by other algorithms that carry their own virtual times. In fact, 
in actual simulations, this messaging phase may take an enormous amount of time, depending 
on the hardware configuration and the message processing algorithms. In our modeling the 
time extent of the messaging phase is ignored as though communications among processors 
were taking place instantaneously. In this sense we model an ideal system of processors. 
The LVT of a cycle represents only the time that logical processes require 
to complete the first phase of a cycle. Therefore, the spread in LVTs represents 
only the desynchronisation that arises due to the asynchronous conservative 
algorithm alone. By the same token, all other performance indicators such as, 
e.g., the overall efficiency or the utilization of the parallel processing 
environment, that are read out of the simulated VTH are the intrinsic properties 
of this algorithm.

This chapter is organized as follows. The simulation model of asynchronous 
conservative updates and the mapping between the logical processes and the 
physical processes considered in this study are explained in Sec.~\ref{model}. 
Section~\ref{physics} outlines the selected ideas taken from non-equilibrium 
surface science that are used in the interpretation of simulation results; 
in particular, the concepts of universality and a non-universal microscopic 
structure that are relevant in deducing algorithmic properties from 
the simulated VTH. One group of these properties includes the utilization 
and the speedup, which is provided in Sec.~\ref{util}. Another group includes 
the desynchronization and the memory request per processor, required for past 
state savings, which is presented in Sec.~\ref{dynamics}. Performance of the 
conservative and the optimistic PDES algorithms is discussed in Sec.~\ref{compare}. 
The new approach to performance studies, outlined in this chapter, can be a very 
convenient design tool in the engineering of algorithms. This issue and directions 
for future research are discussed in Sec.~\ref{conclude}.

\section{Simulated Virtual Time Horizon \label{model}}

In simulations a system of $L$ processors is represented as a set of equally 
spaced lattice points $k$, $k=1, 2,...,L$. Each processor performs a number of 
operations and enters a communication phase to exchange information with its 
immediate neighbors. This communication phase, called an update attempt, takes 
no time in our simulations. In this sense we simulate an ideal system of processors, 
as explained in Sec.~\ref{intro}. An update attempt is assigned an integer index $t$ 
that has the meaning of a wall-clock time (in arbitrary units, which may be thought of 
as a fixed number of ticks of the CPU clock). The local virtual 
time $h_k(t)$ at the $k$-th processor site represents the cumulative local time of 
all operations on the $k$-th processor from the beginning at $t=0$ to time $t$. 
These local processor times are not synchronized by a global clock. 

\begin{figure}[tp]
\unitlength1cm
\includegraphics[width=12.0cm]{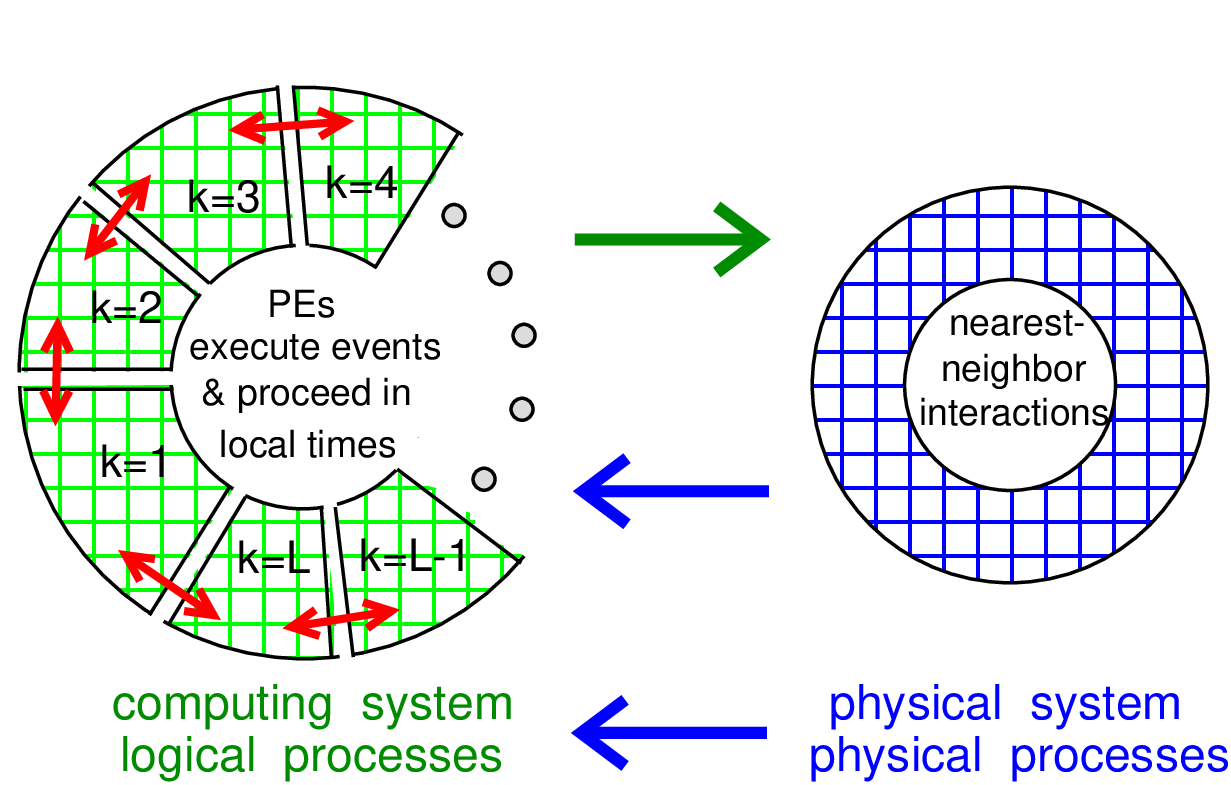}
\caption{\label{kol-01} \baselineskip=0.42truecm
The mapping between physical processes and logical processes.
The nearest-neighbor physical interactions on a lattice with periodic boundary
conditions (the right part) are mapped to the ring communication topology of
logical processes (two-sided arrows in the left part). Each PE carries a
sublattice of $N$ sites. Communications take place only at border
sites. Each PE has at most two effective border sites, i.e., neighboring PEs that 
it communicates with.}
\end{figure}

There is a two-way correspondence between the physical system being simulated in PDES 
and the system of PEs that are to perform these PDES in a manner 
{\it consistent with and faithful to} the underlying stochastic dynamics of the 
physical system, as depicted in Fig.\ref{kol-01}. On the one hand, by spatially 
distributing a physical lattice with the nearest-neighbor interactions and periodic 
boundaries among $L$ processors, the asynchronous nature of physical dynamics is 
carried over to the asynchronous nature of logical processes in the ring 
communication topology of the computing system. On the other hand, the ring 
communication topology among processors is mapped onto a lattice arrangement 
with periodic boundary conditions, $h_{L+k}(t)=h_k(t)$, and asynchronous update 
events in the system of PEs can be modeled as an asynchronous cellular automaton 
on this lattice.

\begin{figure}[tp]
\unitlength1cm
\includegraphics[width=12.0cm]{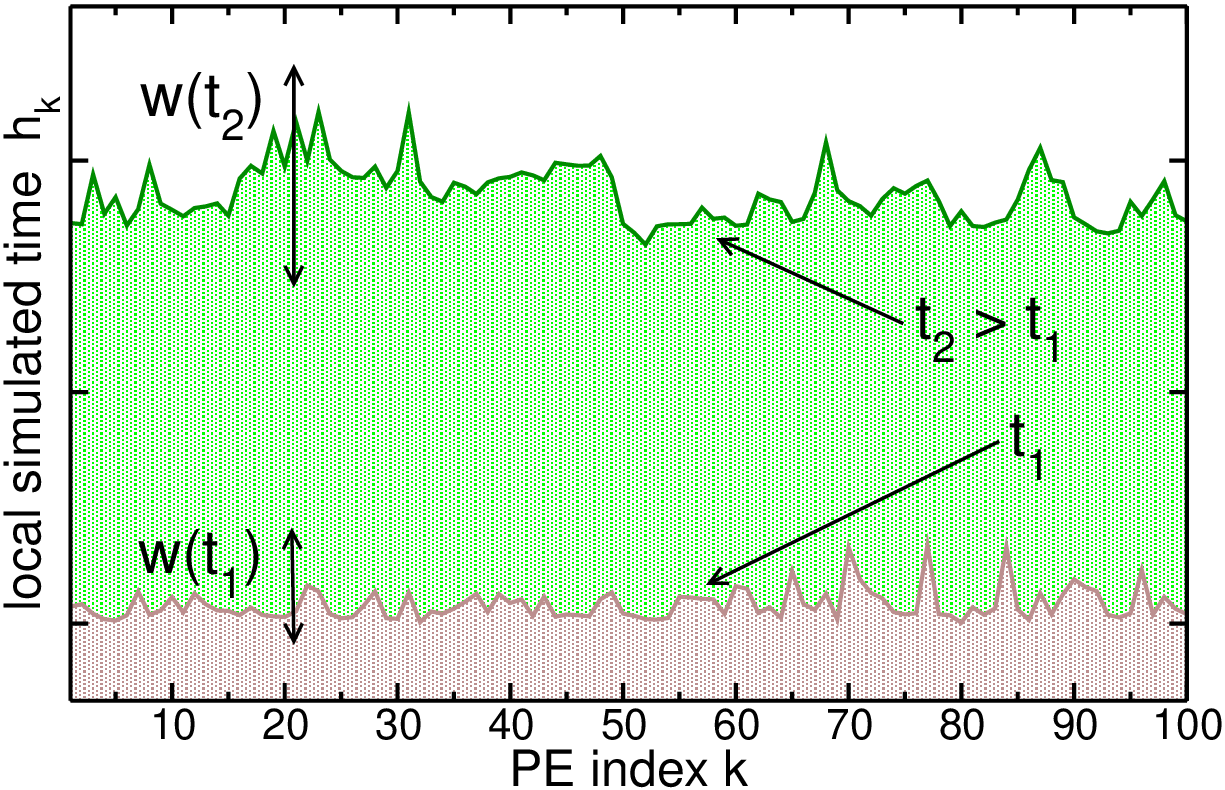}
\caption{\label{kol-02} \baselineskip=0.42truecm
The growth and roughening of the simulated VTH interface: snapshots at $t_1$
and at a later time $t_2$. Local heights $h_k$ are in arbitrary units.
Here, $L=100$ and $N=1$.}
\end{figure}

The set of local virtual times $h_k(t)$ forms the VTH at $t$ (see Fig.~\ref{kol-02}). 
The time-evolution 
of the VTH is simulated by an update rule, where local height increments 
$\eta_k(t)$ are sampled from the Poisson distribution of unit mean $\mu_P=1$. 
The form of the deposition rule depends on the processor load $N$, as explained 
later in this section.

A general principle that governs the conservative update protocol requires a processor 
to idle if at update attempt $t$ the local causality constraint may be violated. 
This happens when at $t$ the $k$-th processor does not receive 
the information from its neighboring processor (or processors) 
if such information is required to proceed in its computation. 
This corresponds to a situation when the local virtual time $h_k(t)$ 
of the $k$-th processor is ahead of either one of the local virtual 
times $h_{k-1}(t)$ or $h_{k+1}(t)$ of its left and right neighbors, 
respectively. In this unsuccessful update attempt the local virtual 
time $h_k(t)$ is not incremented, i.e., the $k$th processor waits: 
$h_k(t+1) = h_k(t)$. In another case, for example, when at $t$ the 
$k$th processor does not need information from its neighbors it performs an update 
regardless of the relation between its local virtual time and the local virtual times on 
neighboring processors. At every successful update attempt, the simulated local virtual 
time at the $k$-th PE-site is incremented for the next update attempt: 
$h_k (t+1) = h_k (t) + \eta_k (t)$,
where $\eta_k(t)= - \ln (r_{kt})$, and $r_{kt} \in (0;1]$ is a uniform random 
deviate. The  simulations start from the {\it flat-substrate condition} at $t=0$: 
$h_k(0)=0$.

One example of computations that follow the above model is a {\it dynamic Monte Carlo} 
simulation for Ising spins \cite{Lub88,LB00,KNR99,NKK03}. In a parallel environment, 
a spin lattice is spatially distributed among $L$ processors in such a way that each 
processor carries an equal load of one contiguous sublattice that consists of $N$ 
spin sites (i.e., each processor has a load of $N$ volumes). 
Some of these $N$ spin-lattice sites belong to border slices, 
i.e., at least one of their immediate neighbors resides on the sublattice of a 
neighboring processor. Processors perform concurrent spin-flip operations 
(i.e., increment their LVTs) as long as a randomly selected 
spin-site is not a border site. If a border spin-site is selected, to perform 
a state update a processor needs to know the current spin-state of the corresponding 
border slice of its neighbor. If this information is not available at the $t$ update 
attempt (because the neighbor's local time is behind), by the conservative update 
rule the processor waits until this information becomes available, i.e., until the 
neighbor's local virtual time catches up with or passes its own local virtual time. 

The least favorable parallelization is when each processor carries the minimal 
load of $N=1$. Computationally, this system can be identified with a closed spin 
chain where each processor carries one spin-site. At each update attempt each 
processor must compare its LVT with the local times on both of 
its neighbors. The second least favorable arrangement is when $N=2$. As before, 
at each update attempt every processor must compare its local time with the local 
time of one of its neighbors. When $N \ge 3$, at update attempt $t$, the comparison 
of the local virtual times between neighbors is required only if the randomly 
selected volume site is from a border slice.

The above three cases are realized in simulations by the following three 
update rules. When $N=1$, the update attempt at $t$ is successful iff 
\begin{equation}
\label{rule1}
h_k (t) \le \min \left\{ h_{k-1} (t), h_{k+1} (t) \right\}.
\end{equation}
When $N=2$, at any site $k$ where the update attempt 
was successful at $(t-1)$, at $t$  we first randomly select a neighbor 
(left or right). This is equivalent to selecting either the left or 
the right border slice on the $k$th processor. The update 
attempt is successful iff 
\begin{equation}
\label{rule2}
h_k (t) \le h_n (t),
\end{equation}
where $n$ is the randomly selected neighbor ($n=k-1$ for the left, 
$n=k+1$ for the right). At any site $k$ where the update attempt 
was not successful at $(t-1)$, at $t$ we keep the last $n$ value. 
When $N \ge 3$, at any site $k$ where the update attempt 
was successful at $(t-1)$, at $t$  we first randomly select any of the 
$N$ volume sites (indexed by $n_k$) assigned to a processor. 
The selected site can be either from the border sites (either $n_k=1$ 
or $n_k=N$) or from the interior. The attempt is successful if the selected site is the 
interior site. When the border site is selected, the attempt is successful if 
condition~(\ref{rule2}) is satisfied. As for $N=2$, at any site $k$ where the 
update attempt was not successful at $(t-1)$, at $t$ we keep the last $n_k$ value. 

In this way the simulated VTH, corresponding to the conservative update rule 
followed by the PDES algorithm, emerges as a one-dimensional non-equilibrium 
surface grown by depositions of Poisson-random time increments that model waiting 
times. Two sample VTH surfaces are presented in Fig.~\ref{kol-02}. Major 
properties of the corresponding algorithm are encoded in these interfaces. 
In principle, with the help of statistical physics, one should be able to 
obtain from VTH such properties as the utilization, the ideal speedup, the desynchronization, 
the memory request per processor, the overall efficiency and the scalability. 
One basic property is the {\it mean utilization} $\langle u(t) \rangle$, which 
can be assessed as the fraction of sites in the VTH interface that performed an 
update at $t$, averaged over many independent simulations. For the minimal load 
per processor $N=1$, $\langle u (t) \rangle$ is simply the mean density of local 
minima of the interface (Fig.~\ref{kol-02}). Another basic property is the 
{\it mean desynchronization} $\langle w(t) \rangle$ in operation times, which 
can be estimated from simulations as the mean statistical spread (roughness) of 
the VTH interface. In the following section we review some useful concepts from 
surface physics relevant to our study.

\section{The $(1+1)$ dimensional nonequilibrium interfaces \label{physics}}

The roughness of a surface that grows on a one dimensional substrate 
of $L$ sites can be expressed by its interface width $w(t)$ at time $t$
\begin{equation}
\label{width}
\langle w^2 (t) \rangle =\left\langle \frac{1}{L} \sum_{k=1}^L
\left( h_k (t) - \bar{h}(t) \right) ^2 \right\rangle,
\end{equation}
where $h_k(t)$ is the height of the column at site $k$ and $\bar{h}(t)$ is 
the average height over $L$ sites, $\bar{h}(t)=(1/L)\sum_{k=1}^L h_k(t)$. 
The angular brackets denote the average over many interface configurations 
that are obtained in many independent simulations. In our 
study these configurational averages were computed over 800 simulations, 
unless noted otherwise. 

Based on the time-evolution of $w(t)$, interfaces can be classified in 
various {\it universality classes} (for an overview see Ref.~\cite{BS95}). 
The idea behind the concept of universality is that, in a statistical description, 
the growth of the surface depends only on the underlying mechanism that generates 
correlations among time-evolving columns $h_k(t)$ and {\it not} on the physical 
particulars of physical (or other) interactions that cause the growth. For instance, 
two completely different physical interactions among deposited constituents 
(e.g., one of a magnetic nature and the other of a social nature) may generate two 
equivalent surfaces of one universality class, depending on the observed evolution 
of $w(t)$. The simplest case of surface growth is {\it random deposition} (RD), 
when the column heights $h_k(t)$ grow independently of each other. The RD interface 
is totally uncorrelated. The time-evolution of its width is characterized by 
a never-ending growth in accordance to the power law $w(t) \sim t^\beta$, 
with the {\it growth exponent} $\beta =1/2$. Such growth defines the RD universality class.

The self-affined roughness of the interface manifests itself by the 
existence of Family-Vicsek scaling \cite{FV85}:
\begin{equation}
\label{Family1}
w^2(t) = L^{2 \alpha} f \left( \frac{t}{L^z}\right),
\end{equation}
where the scaling function $f(y)$ describes two regimes of the width evolution:
\begin{equation}
\label{Family2}
f(y) \sim \left\{ \begin{array} {r@{\quad , \quad}l}
y^{2 \alpha /z} & y \ll 1 \\
\mathrm{const.} & y \gg 1.  \end{array} \right.
\end{equation}
The {\it dynamic exponent} $z$ gives the time-evolution of the lateral correlation 
length $\xi (t) \sim t^{1/z}$, i.e., at a given $t$ the largest distance along the 
substrate between two correlated columns. When $\xi (t)$ exceeds the system size $L$ the 
width saturates and does not grow any more. At saturation, for $t \gg t_{\times}$, 
for a given $L$ the width remains constant and obeys the power law $w \sim L^{\alpha}$, 
where $\alpha$ is the {\it roughness exponent}. The growth phase is the initial phase 
for $t \ll t_{\times}$ before the {\it cross-over time} $t_{\times} \sim L^z$  
at which saturation sets in. The growth phase is characterized 
by the single growth exponent $\beta = \alpha /z$. The roughness, growth and 
dynamic exponents are universal, i.e., their values depend only on the underlying 
mechanism that generates correlations.

A simple continuum model of non-equilibrium growth that leads to scaling is 
provided by the Kardar-Parisi-Zhang (KPZ) equation \cite{KPZ86}:
\begin{equation}
\label{kpz}
h_t = v(t) + \nu \, h_{xx} + \frac{\lambda}{2} \, h_x^2 + \zeta \,  ,
\end{equation}
where $h=h(x,t)$ is the height field (subscripts denote partial derivatives), 
$v(t)$ is the mean interface velocity, $v(t)= \langle d \bar{h}(t)/dt \rangle$, 
and $\zeta (x,t)$ is the uncorrelated Gaussian noise. The coefficients $\nu$ and 
$\lambda$ give the strength of the linear damping and the 
coupling with the nonlinear growth, respectively. A renormalization group 
analysis \cite{BS95,KPZ86} can provide a connection between the stochastic 
growth equation and scaling exponents.

The {\it KPZ universality class}, governed by the dynamics of Eq.(\ref{kpz}), 
is characterized by $\alpha = 1/2$ and $\beta = 1/3$, and the exponent 
identity $\alpha + z =2$. When $\lambda =0$ in Eq.(\ref{kpz}), the growth is 
governed by the linear Edwards-Wilkinson (EW) equation \cite{EW82}. 
The {\it EW universality class} is characterized by $\alpha = 1/2$ and 
$\beta = 1/4$, and the EW exponent identity is $2 \alpha + 1 = z$. 
When $\lambda =0$ and $\nu =0$ in Eq.(\ref{kpz}), the growth belongs to 
the RD universality class. Unlike the KPZ and EW interfaces, the RD interface 
is not self-affined.

The origins of scale invariance, as in Eq.(\ref{Family1}), and the universal 
properties of time-evolving surfaces are well understood \cite{BS95}. 
In this study, we use the universal properties of the simulated VTH to 
investigate the scalability of the corresponding PDES algorithm. However, 
there are many instances where non-universal properties, i.e., those pertaining 
to the {\it microscopic structure} of the interface, are of importance. 
In this study, one example is the density of local minima or the density 
of update sites of the VTH interface. It is safe to say that there is no 
general silver-bullet-type of approach to these problems. For the case study 
of VTH interfaces, we were able to develop a discrete-event analytic technique 
that provides a means for calculating a probability distribution for events 
that take place on the surface \cite{KNR03,KN04}. For the closed linear chain 
of $L$ processors carrying minimal load, when the VTH is simulated by Poisson-random 
depositions at local interface minima, the probability distribution $P(p;L)$ of 
the update events on the corresponding VTH surface is \cite{KNR03}:
\begin{equation}
\label{distr}
P(p;L) = \frac{1}{2^{L-2}} {L-1 \choose 2p-1},
\end{equation}
where $p$ is the number of updates at the $t$-th update attempt after the simulations 
reach a  steady state. This distribution can be used to derive approximate closed 
formulas for mean quantities measured in simulations, e.g., for $\langle u(t) \rangle$. 
Also, it is a starting point in the derivation of analogous distributions $P(p;L; N)$ 
for the cases when each processor carries a larger load of $N=2$ or $N \ge 3$ \cite{KN04}. 
The advantage of knowing $P(p;L; N)$ is that it enables one to compute analytically 
quantities that otherwise can be only estimated qualitatively in simulations.

\section{Utilization of the Parallel Processing Environment \label{util}}

\begin{figure}[tp]
\unitlength1cm
\includegraphics[width=10.0cm]{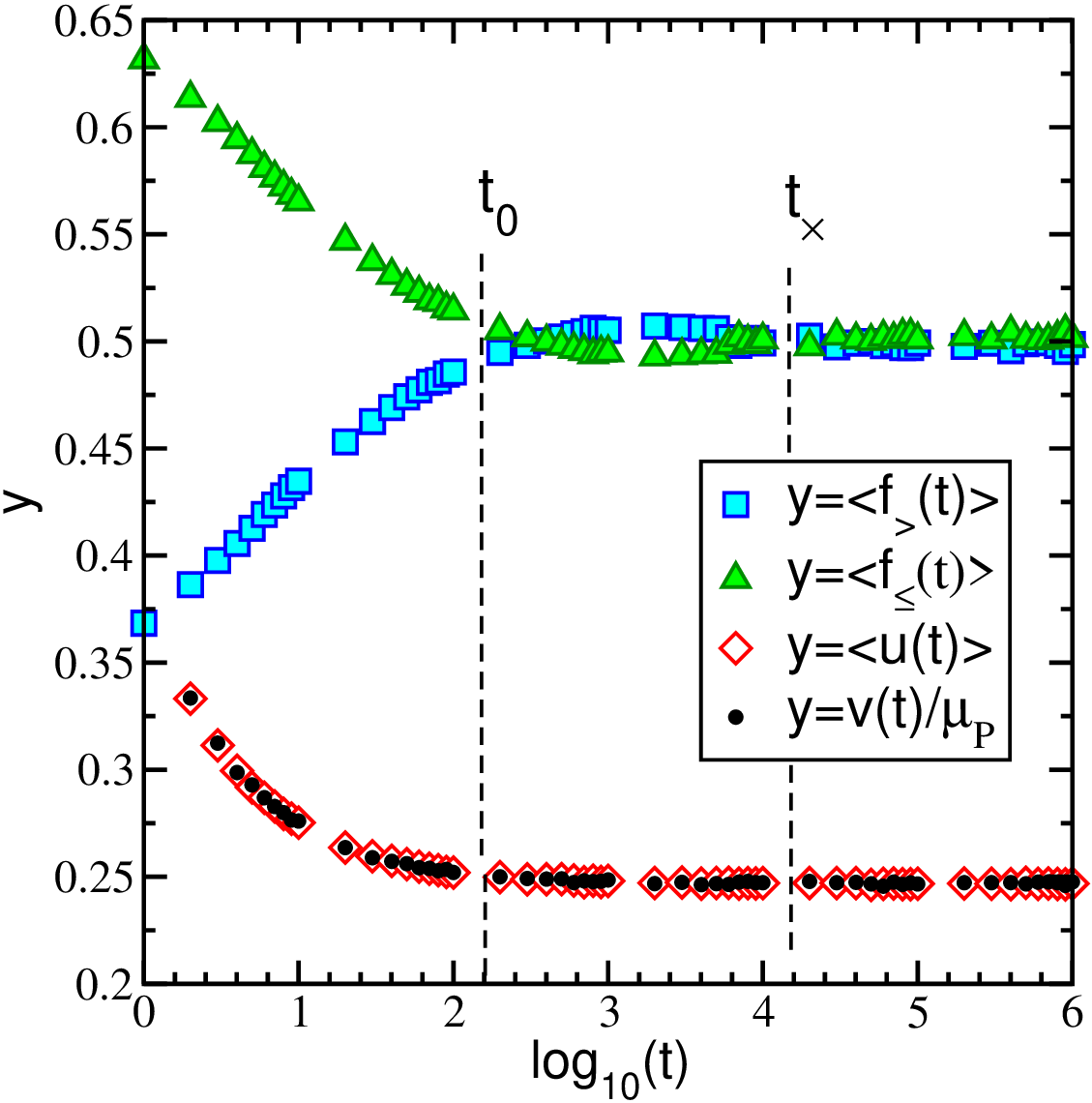}
\caption{\label{kol-03} \baselineskip=0.42truecm
Simulated time evolutions of characteristic densities and the scaled VTH 
interface velocity during simulations with the minimal load $N=1$ per 
processor ($L=1000$). Time $t_0$ marks the transition to steady-state 
simulations and $t_{\times}$ is the saturation time, as explained 
in the text. For times later than $t_0$, both the utilization 
$\langle u \rangle$ (diamonds) and the velocity $v$ (filled circles) are constant.}
\end{figure}
\begin{figure}[tp!]
\includegraphics[width=10.0cm]{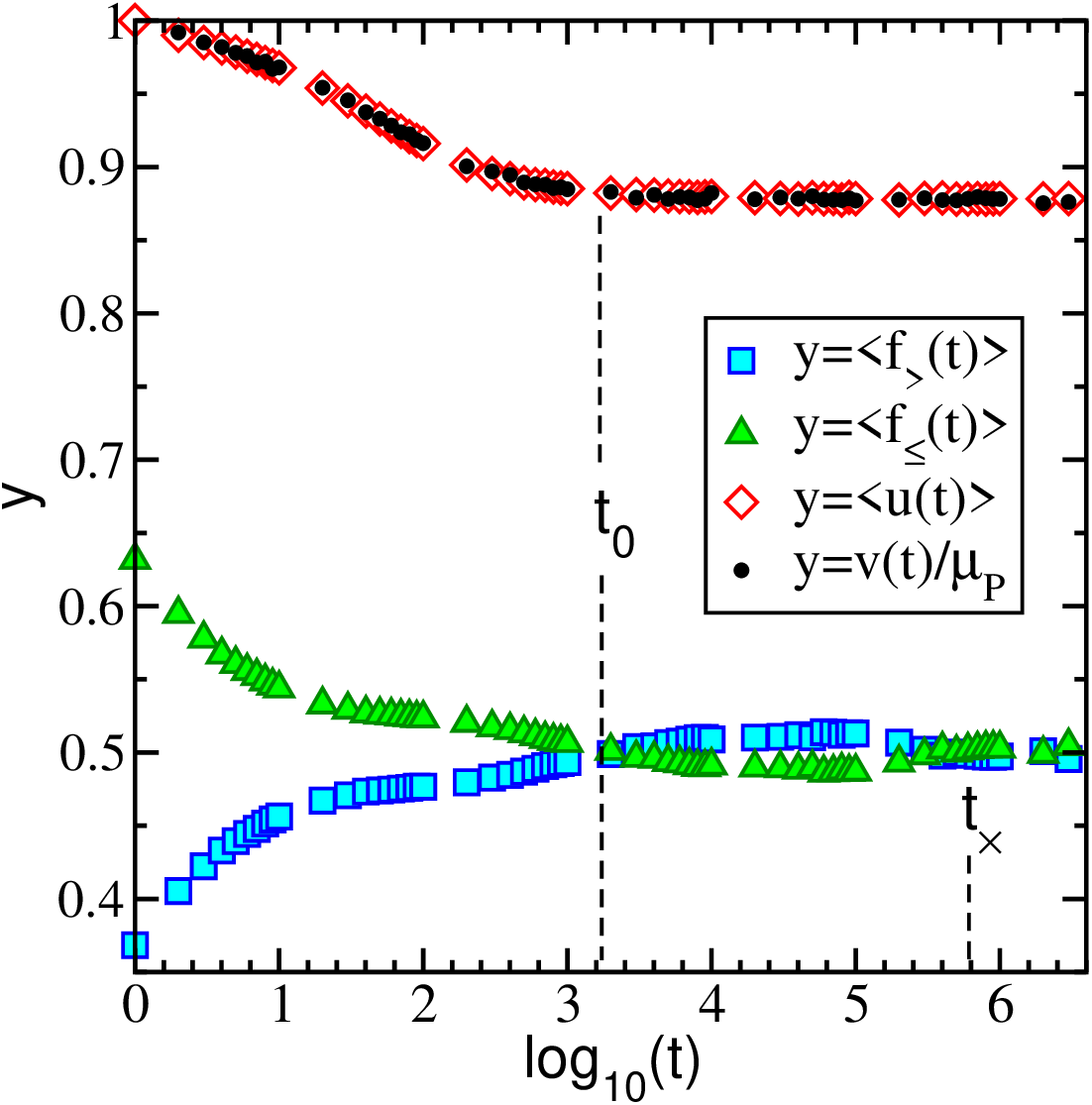}
\caption{\label{kol-04} \baselineskip=0.42truecm
Simulated time evolutions of characteristic densities and the scaled 
VTH interface velocity during simulations with the load $N=100$ per 
processor ($L=1000$). As in Fig.~\ref{kol-03}, for times later 
than $t_0$ both the utilization $\langle u \rangle$ (diamonds) and 
the velocity $v$ (filled circles) are constant, but their values are 
significantly higher than in the worst-case performance scenario 
presented in Fig.~\ref{kol-03}.}
\end{figure}

For a real system, the mean utilization $\langle u(t) \rangle$ is defined either
as the number or as the fraction of processors that on average work simultaneously
at a time. In our model, $\langle u(t) \rangle$ is the mean fraction (i.e., the density)
of update sites in the simulated VTH. The simulated time evolution of $\langle u(t) \rangle$
is presented in Fig.~\ref{kol-03} (for $N=1$) and in Fig.~\ref{kol-04} (for $N=100$)
that illustrate the following observations. First, $\langle u(t) \rangle$ is not
constant as the simulations evolve but abruptly decreases from its initial value at
$t=0$ and very quickly, after a few hundred steps, settles down at its steady value
when it no longer depends on time. Both the utilization and the transition period
$t_0$ to the {\it steady state} depend strongly on the processor load $N$. Second,
the VTH velocity $v(t)$ must be related to $\langle u(t) \rangle$ by a simple linear
scaling relation. Third, the transition time $t_0$ to the steady state can be
estimated from simple statistics of the interface. Let us elaborate on these issues.

The characteristic densities $\langle f_{>}(t) \rangle$ and $\langle f_{\le}(t) \rangle$, 
plotted in Figs.~\ref{kol-03}--\ref{kol-04}, are the fractions of the interface sites 
(processors) that have their LVT larger and equal-or-smaller, respectively, 
than the mean virtual time $\bar{h}(t)$. Their relation to each other is a simple 
indicator of the skewness of the distribution of the LVTs about the mean virtual time. 
For the times when they approximately coincide this distribution is approximately 
symmetric. The reason for a non-zero skewness for the early times $t<t_0$ is the 
flat-substrate initial condition, i.e., the initial null LVT on all processors 
(the detailed analysis of this issue can be found in \cite{KNV04}). In the worst-case 
performance scenario, when each processor has the minimal load of $N=1$ 
(Fig.~\ref{kol-03}), the duration of this initial transition time $t_0$ to the 
steady state is a non-universal property of the VTH interface. This means that 
in real applications the time $t_0$ will depend on the application platform, 
i.e., the hardware configuration and parameters. But, for a real application $t_0$ 
can be determined in a non-expensive way by monitoring $\langle f_{>}(t) \rangle$ 
and $\langle f_{\le}(t) \rangle$ in a trial simulation with a fixed $N=N_0$. 
Then, the results can be scaled either up or down for an arbitrary load $N$. 
The existence of such scaling is the universal property of the VTH interface, 
which is discussed in Sec.\ref{dynamics} where the explicit scaling relations are given.

\begin{figure}[tp]
\unitlength1cm
\includegraphics[width=10.0cm]{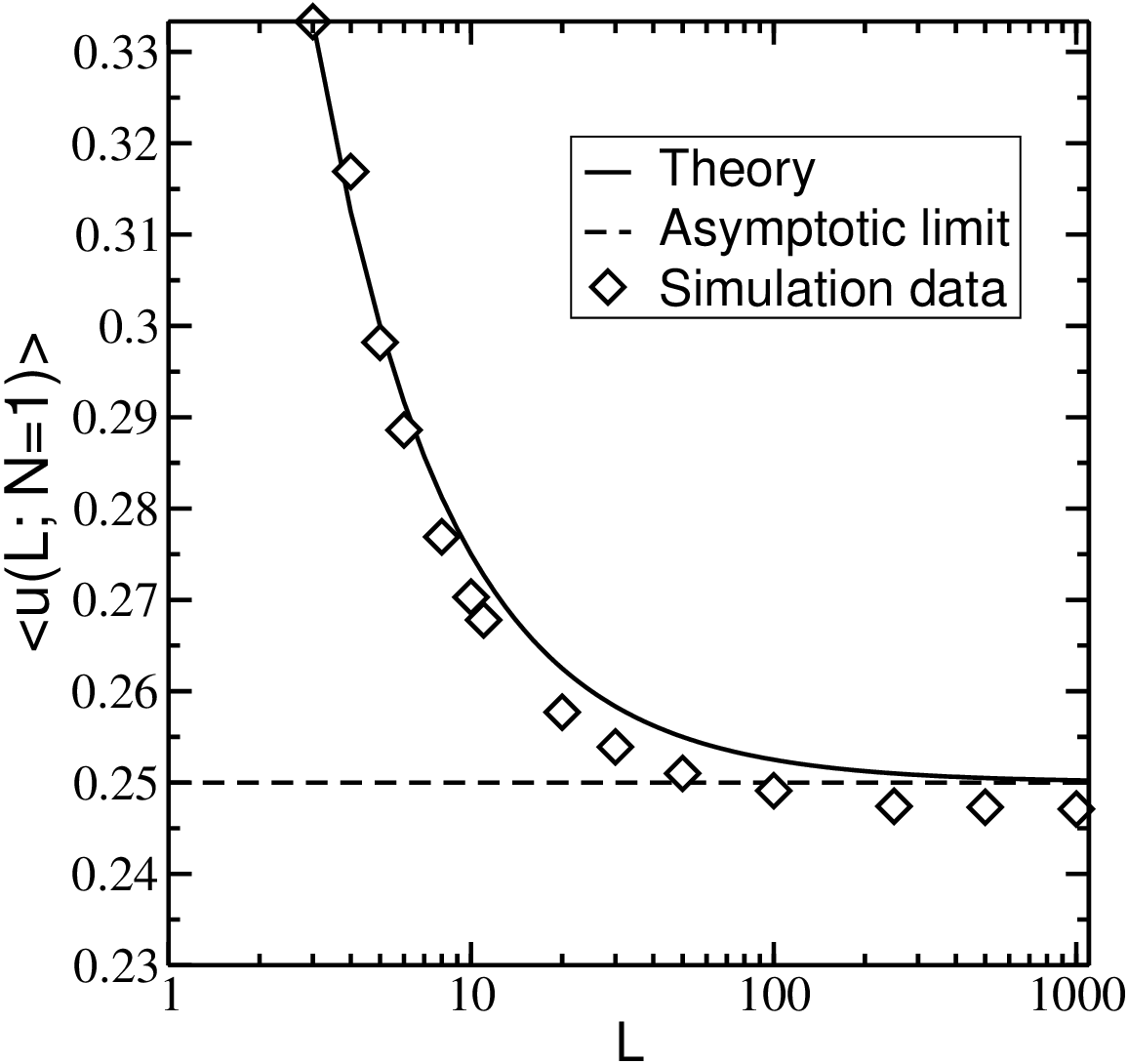}
\caption{\label{kol-05} \baselineskip=0.42truecm
The steady-state mean utilization vs the number of processors $L$ for
the minimal load per processor: the analytical result (continuous curve),
its asymptotic limit $\lim_{L \to \infty} \langle u(L;1) \rangle =1/4$
(horizontal line) and simulation results (symbols). The error bars are
smaller than the symbol size.}
\end{figure}

The overall progress of PDES can be estimated from the time rate of the 
{\it global virtual time} (GVT) that is the smallest LVT from all processors at $t$. 
The GVT determines the fossil collection, i.e., the memory that can be re/de-allocated 
from past-saved events. In our model, the mean GVT is the mean global minimum 
$\langle h_{GVT}(t) \rangle$ of the simulated VTH: 
$h_{GVT}(t) = \min\{h_k(t): 1 \le k \le L\}$. On the average, the time rate of 
$\langle h_{GVT}(t) \rangle$ is not larger than the VTH interface velocity $v(t)$ 
and after cross-over time $t_{\times}$ to saturation (indicated in 
Figs.~\ref{kol-03}--\ref{kol-04} and Figs.~\ref{kol-08}--\ref{kol-10}) these 
two rates are equal. Thus, $v(t)$ is the measure of progress. It is shown 
analytically \cite{KNV04} that $v(t)=\langle u(t) \rangle \mu$, where $\mu$ 
is a constant. For the 
simulated VTH $\mu = \mu_P =1$, as can be seen in Figs.~\ref{kol-03}--\ref{kol-04}. 
In a real application $\mu$ is a hardware dependent parameter. 

\begin{figure}[tp!]
\includegraphics[width=10.0cm]{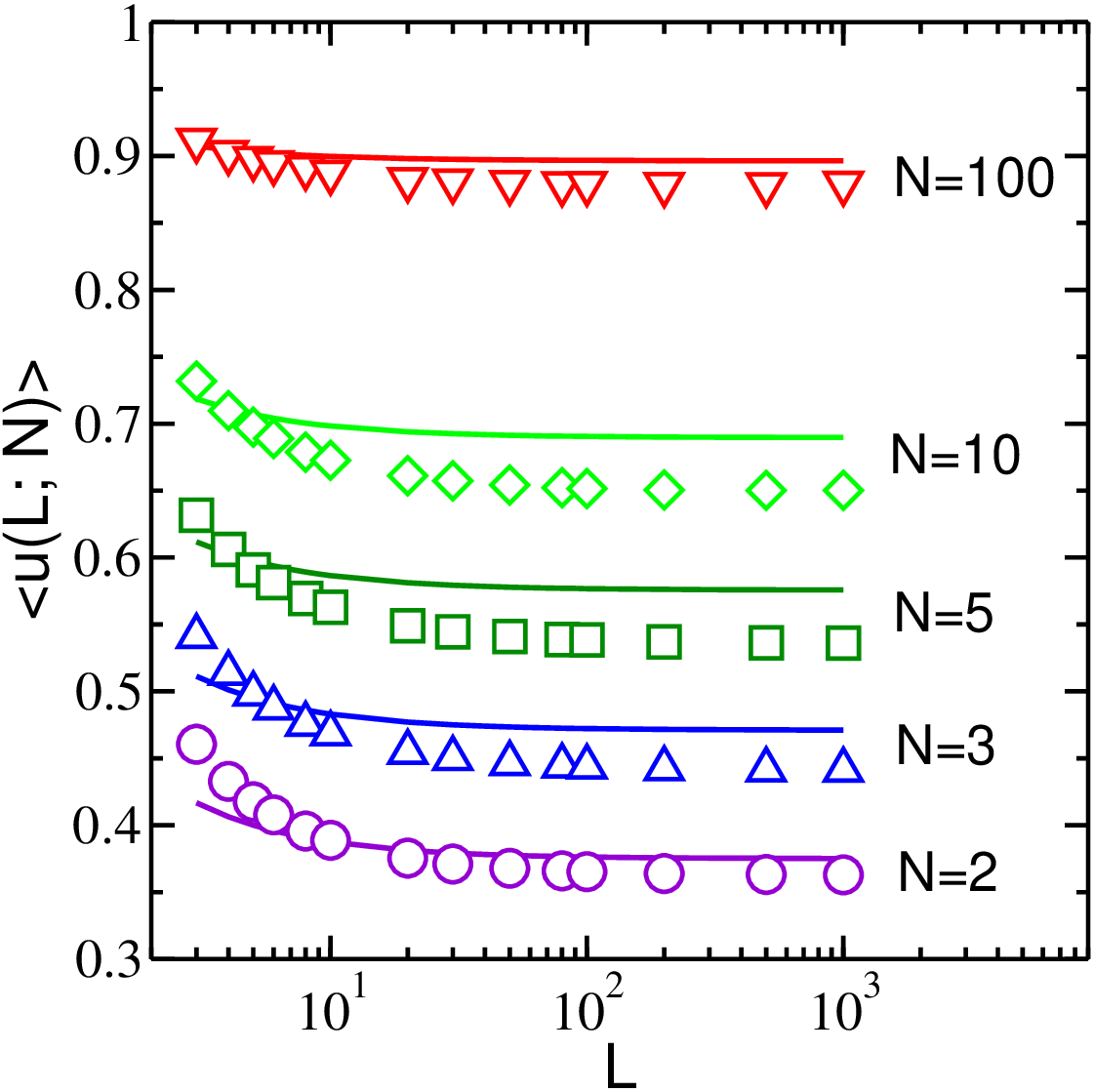}
\caption{\label{kol-06} \baselineskip=0.42truecm
The steady-state mean utilization vs the number of processors $L$ for
loads $N \ge 2$ per processor: analytical results (continuous curves) and
simulation results (symbols). The error bars are smaller than the symbol size.}
\end{figure}

Since the overall progress in PDES connects linearly to $\langle u(t) \rangle$, 
the mean utilization is the most important property of this algorithm. 
The steady-state utilization $\langle u(L; N=1) \rangle$ for the worst-case 
performance as a function of the number $L$ of processors is presented in 
Fig.~\ref{kol-05}. For brevity of notation, from now on we omit the index 
$t$ because for $t>t_0$ the utilization is constant. Considering the simulation 
data alone (symbols in Fig.~\ref{kol-05}) it is easily seen that even at the infinite 
limit of $L$ the mean utilization has a non-zero value. In fact, this asymptotic 
value is approached very closely with less than a thousand  processors. The existence 
of this limit guarantees a non-zero progress of PDES for any value of $L$ and for 
any processor load $N$, since the curve $\langle u(L; N=1) \rangle$ is the lower 
bound for the steady-state utilization $\langle u(L; N) \rangle$ 
(compare with Fig.~\ref{kol-06}).

The presence of this non-zero limit for $\langle u(L; 1) \rangle$ as
$L \to \infty$ and the global behavior of the simulation data, observed
in Fig.~\ref{kol-05}, suggest the existence of some underlying scaling law
for $\langle u(L; 1) \rangle$. It appears that, indeed, the underlying
scaling law can be obtained analytically from the first moment of the probability
distribution $P(p;L)$ given by Eq.(\ref{distr}) and the result is \cite{KNR03}
\begin{equation}
\label{util-1}
\langle u(L \ge 3; 1) \rangle = \frac{L+1}{4 L},
\end{equation}
and $\langle u(L=2; 1) \rangle =1/2$. This function is plotted in Fig.~\ref{kol-05}
for $L \ge 3$. Similarly for any $N \ge 2$, starting with $P(p;L)$, one can construct
the probability distribution $P(p;L; N)$ of updates in the system of $L$ processors,
each having a load of $N$ (i.e., the distribution of updates on the corresponding
VTH interface). The mean utilization $\langle u(L;N) \rangle$ can then be obtained
from the first moment of $P(p;L; N)$ \cite{KN04}:
\begin{equation}
\label{util-2}
\langle u(L \ge 3; N \ge 2) \rangle = \left( 1- \frac{q(N)}{2} \right) 
\left( 1- \frac{q(N)}{4} \, \frac{L-1}{L} \right), 
\end{equation}
where $q(N)= \sqrt{2/N}$, and
\begin{equation}
\label{util-3}
\langle u(L=2; N \ge 3) \rangle = 1- \frac{1}{\sqrt{2N}} ,
\end{equation}
and $\langle u(L=2; N =2) \rangle =1/2$. Relation~(\ref{util-3}) is exact. 
Relation~(\ref{util-2}) is 
presented in Fig.~\ref{kol-06}. As $L \to \infty$ the asymptotic limit is 
$\langle u(\infty; N \ge 3) \rangle = (2-q)(4-q)/8 > \langle u(\infty; N=1) 
\rangle = 1/4$.

The computational speedup $s$ of a parallel algorithm is defined as the ratio 
of the time required to perform a computation in serial processing to the time 
the same computation takes in concurrent processing on $L$ processors. 
It is easy to derive from the above definition that for an {\it ideal system 
of processors}, that is for the particular update {\it algorithm}  
considered in this work, the mean speedup is
\begin{equation}
\label{speedup-1}
\langle s \rangle = L \langle u(L;N) \rangle . 
\end{equation}
In other words, $\langle s \rangle$ is measured by the average number of PEs that 
work concurrently between two successive update attempts.

We observe that for ideal PEs the speedup as a function $F(L)$ must be such 
that the equation $F(L)=s$ has a unique solution, where $s$ is a fixed positive 
number. This requirement follows naturally from the logical argument that 
distributing the computations over $L$ ideal PEs gives a unique speedup, i.e., 
two ideal systems having sizes $L_1$ and $L_2$, respectively, may not give 
the same $s$. This means that $F(L)$ must be a monotonically increasing function of $L$.

Combining Eq.(\ref{speedup-1}) and Eq.(\ref{util-1}), in the worst-case performance 
when the load per processor is minimal, the mean speedup is 
$\langle s(L;N=1) \rangle = (L+1)/4$ and $\langle s(L=2,3;N=1) \rangle = 1$. 
The latter relation says that this algorithm produces no speedup when the 
computations are distributed over $2$ or $3$ processors. In this case, although 
the utilization is $1/2$, the processors do not work concurrently but 
{\it alternately}, i.e., one is working while the other is idling. 
For a real system of PEs performing PDES, in such a situation the communication 
overhead will produce an actual slowdown, i.e., the parallel execution time will 
be longer than the sequential execution time on one processor. When $L \ge 4$, 
to take an advantage of concurrent processing the average number of PEs working 
in parallel between two successive update attempts must satisfy $(L+1)/4 > 2$, 
which gives $L>7$. Still, depending on the implementation platform, the actual 
speedup of PDES may be negligible or not present at all for small $L$.

For a general load per processor, combining Eq.(\ref{speedup-1}) and 
Eq.(\ref{util-2}) gives a linear relation with respect to $L$:
\begin{equation}
\label{speedup-2}
\langle s (L \ge 3; N \ge 2) \rangle =  
\left[ L \left( 1- \frac{q(N)}{4} \right) + \frac{q(N)}{4} \right] 
\left( 1 - \frac{q(N)}{2} \right).
\end{equation}
Equation~(\ref{speedup-2}) can be rearranged to 
\begin{equation}
\label{speedup-3}
\langle s (L \ge 3; N \ge 2) \rangle = \frac{L-1}{8} 
\left( \bar{q}(N) + \frac{2L}{L-1} \right)^2 - \frac{(L+1)^2}{8(L-1)}, 
\end{equation}
where $\bar{q}(N)=1-q(N)$ is the probability that a processor performs 
an update without the  need of communicating with other processors 
\cite{KN04,KNV04}. This probability sharply increases with the processor load. 
Since the mean speedup increases quadratically with $\bar{q}(N)$ and only 
linearly with $L$, for some PDES that perform the updates in accordance to this 
algorithm it may be more advantageous to assign more load per processor than 
to distribute the computations over a large number of processors. In any case, 
either of the above relations,  Eq.(\ref{speedup-2}) or~(\ref{speedup-3}), can be 
used to assess the upper bound for the speedup in actual applications.

\section{Dynamics of Desynchronization \label{dynamics}}

\begin{figure}[tp]
\includegraphics[width=10.0cm]{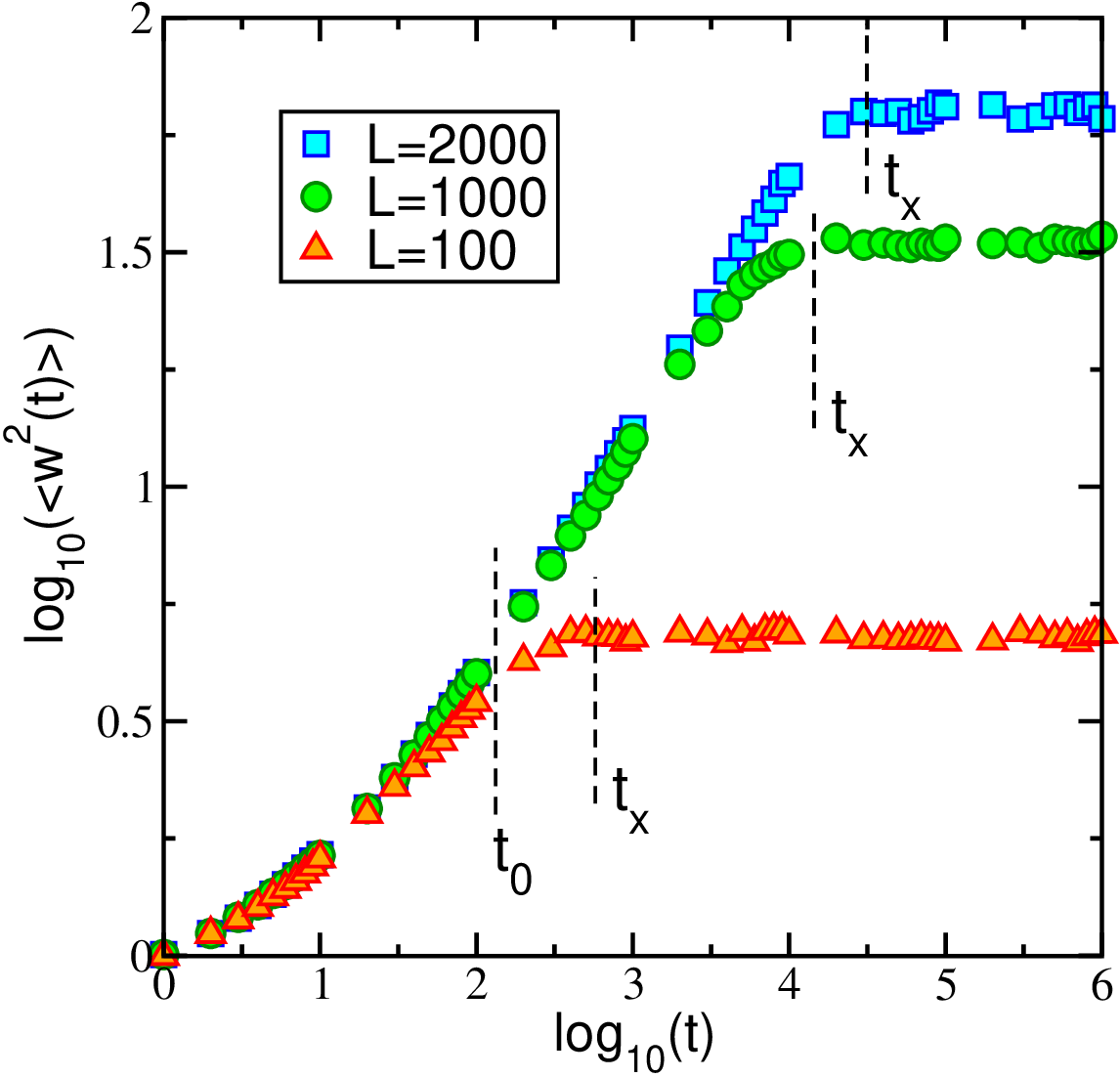}
\caption{\label{kol-07} \baselineskip=0.42truecm
Simulated time evolution of the VTH width in the worst-case performance
scenario $N=1$. Time $t_0$, common for
all $L$, marks the transition to the steady-state simulations of Fig.~\ref{kol-03}.}
\end{figure}
\begin{figure}[tp!]
\includegraphics[width=10.0cm]{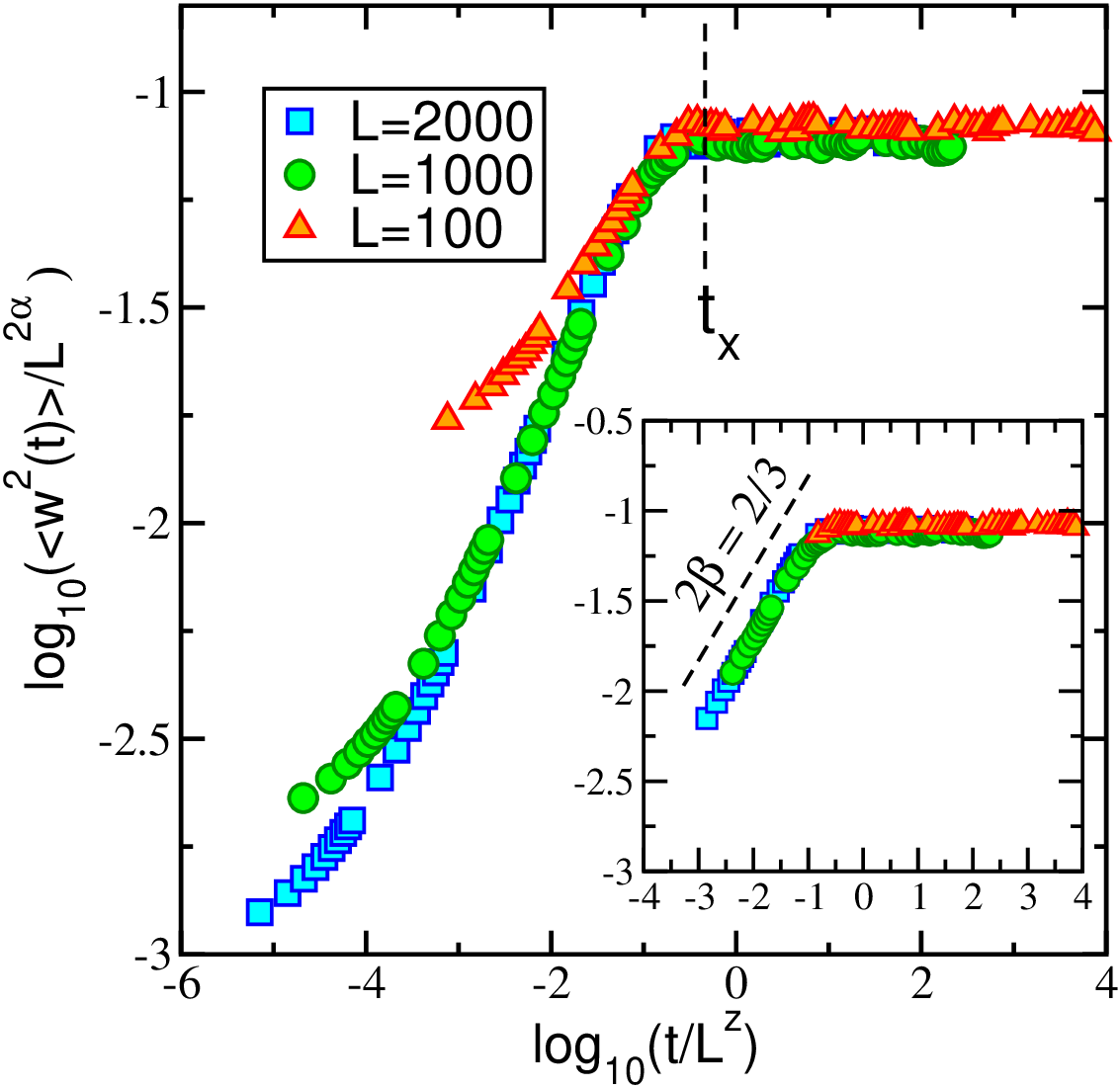}
\caption{\label{kol-08} \baselineskip=0.42truecm
The scaled time evolution of the simulated VTH widths of Fig.~\ref{kol-07}
for all times $t>0$. The insert shows the full data collapse for $t>t_0$,
with the growth exponent $\beta = 1/3$.}
\end{figure}

In PDES the memory request per processor, required for past-state savings, 
depends on the extent to which processors get desynchronized during simulations. 
In our model, the statistical spread $w(t)$ of the simulated VTH, as illustrated 
in Fig.~\ref{kol-02}, provides the measure of this desynchronization. 
In simulations the width of the VTH interface is computed using  
Eq.~(\ref{width}). The representative results of simulations for the case of 
the minimal load per processor are presented in Fig.~\ref{kol-07}. For any 
number $L$ of processors the time evolution of the VTH width has two phases. 
The first phase, for $t \ll t_{\times}$, is the growth regime, where for $t>t_0$ 
the width $w(t)$ follows a power law in $t$ with the growth exponent $\beta = 1/3$. 
The second phase, after the cross-over time $t_{\times}$, is the saturation regime, 
where $w(t)$ has a constant value that depends only on the system size and follows 
a power law in $L$ with the roughness exponent $\alpha = 1/2$. The values of these 
exponents are characteristic of the KPZ universality class. Explicitly, the 
evolution can be written as 
\begin{equation}
\label{evol-1}
\langle w^2 (t) \rangle \sim \left\{ \begin{array} {r@{\quad , \quad}l}
t^{2 \beta} & t_0  \ll t \ll t_{\times} \\
L^{2\alpha} & t \gg t_{\times} ,  \end{array} \right.
\end{equation}
where $t_0$ is the initial regime where the Family-Vicsek scaling law, 
Eqs.(\ref{Family1}--\ref{Family2}), does not hold. This can be seen 
directly when the scaling is performed for all $t>0$, as  
illustrated in Fig.~\ref{kol-08}. The whisker-like structures that appear 
after data collapse in the growth phase, clearly observed in Fig.~\ref{kol-08}, 
indicate that in the initial start-up time $t<t_0$ the curves in Fig.~\ref{kol-07} 
before scaling follow one evolution for all $L$. The insert shows the universal 
Family-Vicsek scaling function, Eq.(\ref{Family2}), for 
$t>t_0$. Here, the cross-over time scales as $t_{\times} \sim L^z$, where 
$z=\alpha/\beta =3/2$. The presence of the initial non-scaling growth regime 
is an artifact of the flat-substrate initial condition. Its duration $t_0$ 
is a non-universal parameter that can be determined in PDES by monitoring 
characteristic densities and the utilization, as discussed in Sec.~\ref{util}.

\begin{figure}[tp]
\includegraphics[width=10.0cm]{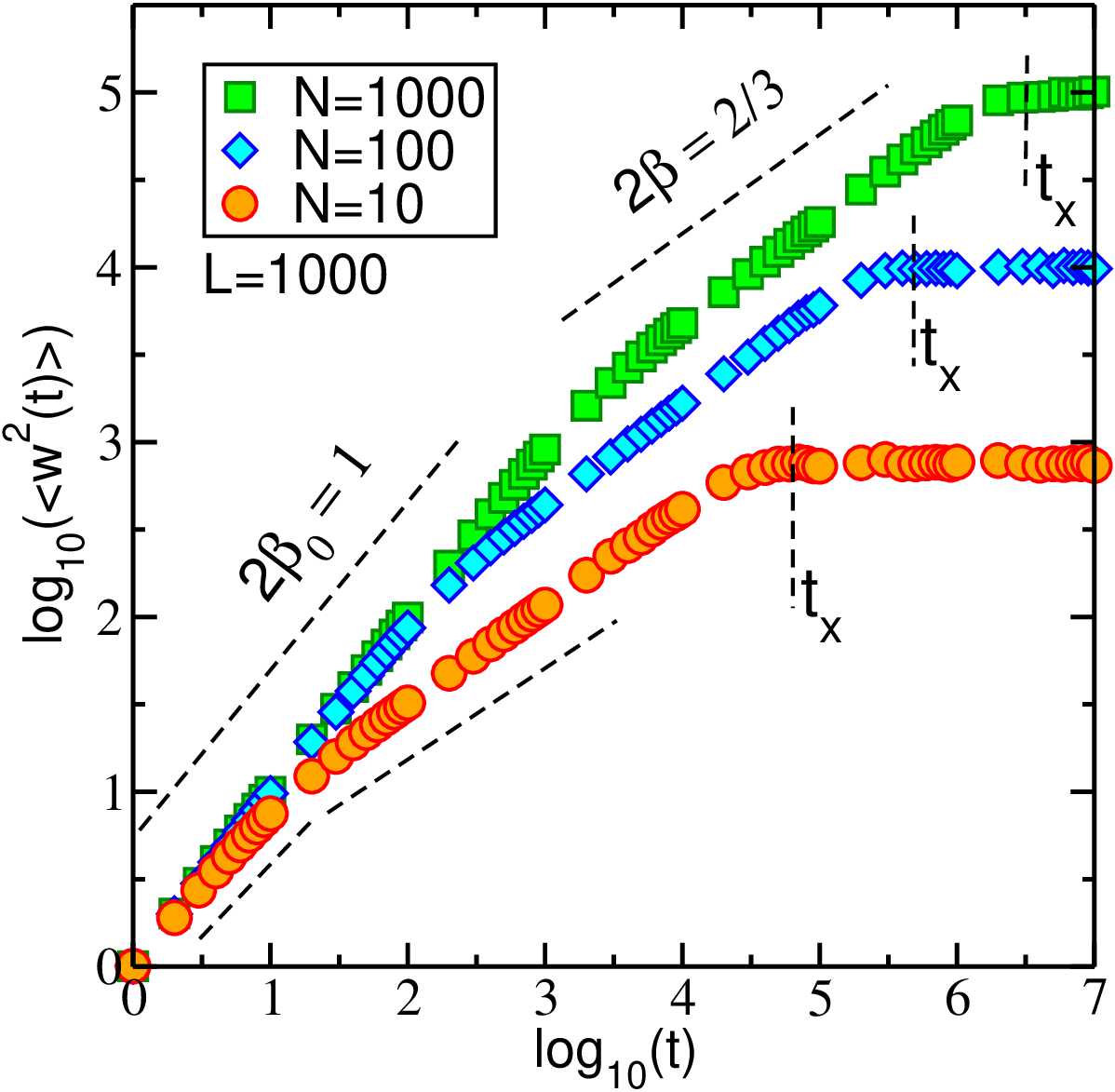}
\caption{\label{kol-09} \baselineskip=0.42truecm
Simulated time evolution of the VTH width for loads  
$N>1$. There are two growth regimes, characterized by two exponents 
$\beta_1$ and $\beta_2$. The duration of the early phase depends on $N$. 
In this early phase, simulations are not in the steady-state: the squared 
width increases linearly with time.}
\end{figure}
\begin{figure}[tp!]
\includegraphics[width=10.0cm]{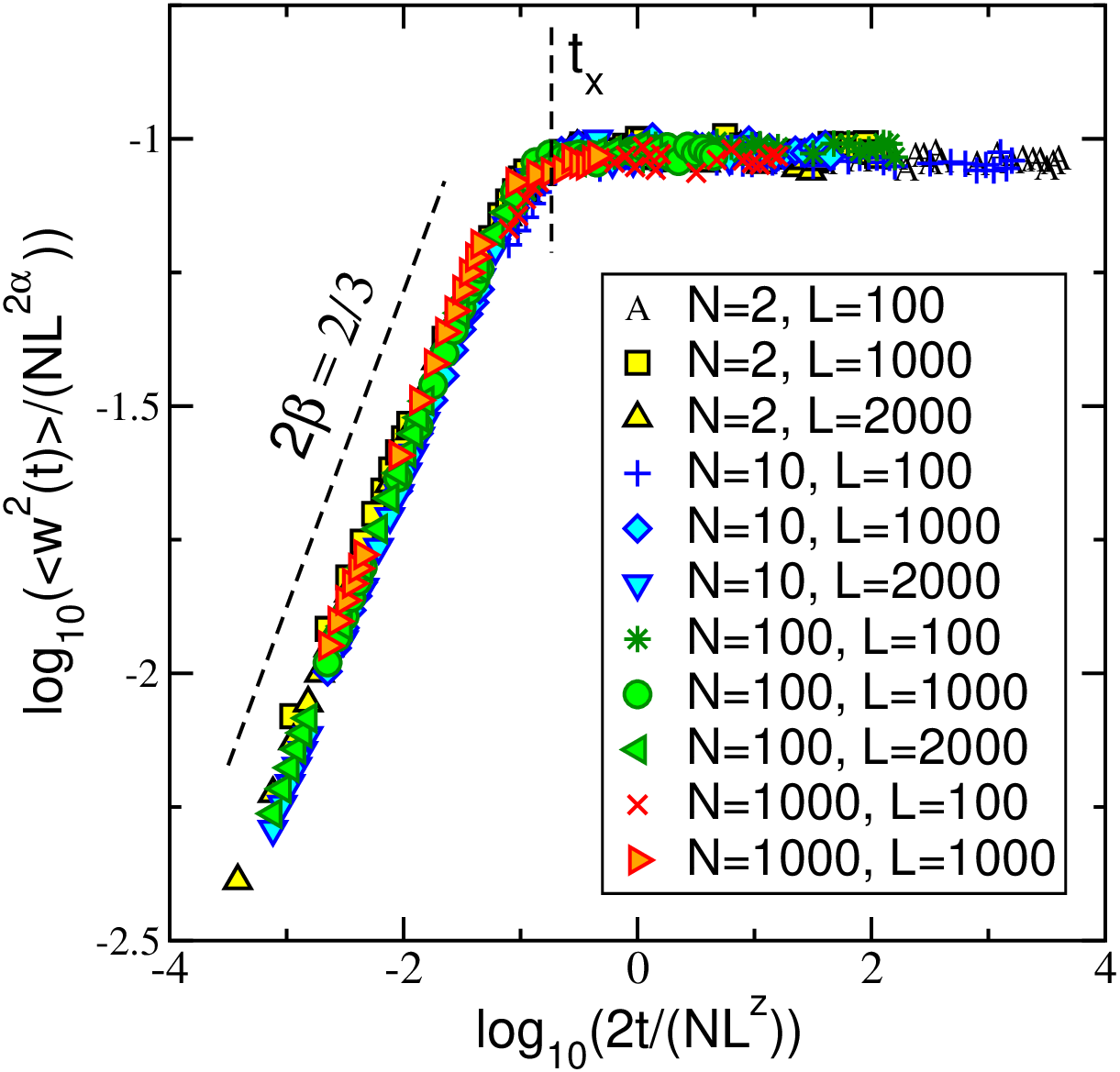}
\caption{\label{kol-10} \baselineskip=0.42truecm
The scaled time evolution of the simulated VTH widths for general values of $N$ 
and $L$. This scaling function is valid only for steady-state simulations. 
In the scaling regime the growth exponent is $\beta = 1/3$, as in Fig.~\ref{kol-08}. 
The time to saturation, when the width is constant, depends on both $N$ and $L$.}
\end{figure}

When the simulations are performed for the case when each processor carries 
a load $N$, the evolution of the VTH width changes. Now, as illustrated in 
Fig.~\ref{kol-09} for $L=1000$, there are two distinct phases in the initial 
growth regime. The early phase evolves in the RD fashion, having the growth 
exponent $\beta_0=1/2$, and the later phase has signatures of the KPZ scaling:  
\begin{equation}
\label{evol-2}
\langle w^2 (t) \rangle \sim \left\{ \begin{array} {r@{\quad , \quad}l}
t^{2 \beta_0} & t < t_0 (N) \\
t^{2 \beta} & t_0 (N) \ll t \ll t_{\times} (N) \\
L^{2\alpha} & t \gg t_{\times}(N) ,  \end{array} \right.
\end{equation}
where both $t_0$ and $t_{\times}$ depend on the processor load. 
The initial RD-like growth does not scale with $L$. This lack of scaling 
extends to approximately $t_0 (N) \propto t_0 N/2$, where $t_0$ marks the end 
of the initial relaxation period when $N=1$ or $N=2$. The physical justification 
for the presence of the RD growth is that when $N \ge 3$ there is a non-zero 
probability of having some processors performing state updates without the need of 
communicating with other processors. This probability of ``uncorrelated" updates 
increases when the processor load increases. However, even for a large but finite 
$N$ there are some processors that may not complete an update without communication 
with another processor, thus, correlations are build among the processors and 
propagate throughout the system. Eventually these ``correlated" updates cause 
the VTH interface to saturate. The net effect of having a large load per processor 
is the noticeable elongation of the time scale over which the correlations are build, 
but the dynamics of building these correlations belongs to the same universality 
class as in the case of the minimal load per processor. Therefore, it is expected 
that the simulated VTH should exhibit KPZ universality in this case as well, 
as soon as the correlations become apparent. Indeed, after the initial transition 
time $t_0(N)$ the VTH widths can be collapsed onto the following scaling 
function \cite{KNV04}:
\begin{equation}
\label{evol-3}
\langle w^2(t) \rangle = \frac{N}{2} L^{2\alpha} f \left(\frac{2}{N}\frac{t}{L^z}\right),
\end{equation}
where $f(y)$ satisfies Eq.(\ref{Family2}), and $z=2-\alpha$ with $\alpha \approx 1/2$. 
This scaling function is presented in Fig.~\ref{kol-10}. Accordingly, the VTH 
interfaces belong to 
the KPZ universality class. In the scaling regime, the evolution can be written 
explicitly as
\begin{equation}
\label{evol-4}
\langle w (t) \rangle \sim \left\{ \begin{array} {l@{\quad , \quad}l}
\left( \sqrt{N/2} \right)^{1-2\beta} \, t^{ \beta} & t_0 (N) \ll t \ll t_{\times} (N) \\
\sqrt{N/2} \, L^{\alpha} & t \gg t_{\times}(N) ,  \end{array} \right.
\end{equation}
where $t_{\times}(N) \sim (N/2) L^z$, $\beta=1/3$, $\alpha=1/2$ and $z=3/2$. 
For $0 \le t<t_0(N) \sim t_0 N/2$, for all $L$ the width follows the power low 
$\langle w(t) \rangle \sim t^{\beta_0}$, where $\beta_0=1/2$.

The consequence of Eq.(\ref{evol-4}) is that the memory request per processor 
does not grow without limit but varies as the computations evolve. The fastest 
growth characterizes the initial start-up phase. The length of the start-up phase 
depends on the load per processor. The start-up phase is characterized by decreasing 
values of the utilization. In the steady-state simulations, when the utilization 
has already stabilized at a mean constant value the memory request grows slower, 
at a decreasing rate $\sim 1/t^{2/3}$. In this phase, the mean request can be 
estimated globally from Eq.(\ref{evol-4}). The important consequence of scaling, 
expressed by Eq.(\ref{evol-3}), is the existence of the upper bound for the 
memory request for any finite number of processors and for any load per 
processor. On the average, this upper bound increases proportionally to 
$\sqrt{NL}$ with the size of conservative PDES.

The characteristic time scale $t_0(N)$ from the first step to the steady-state 
simulations can be estimated by monitoring the utilization for the minimal 
processor load (to determine $t_0$) and, subsequently, scaling this time with $N$. 
Similarly, the characteristic time scale to $t_\times(N)$, when the desynchronization 
reaches its steady state, can be scaled with the processor load to determine 
an approximate number of simulation steps to the point when the mean memory request 
does not grow anymore.

\section{Conservative vs Optimistic \label{compare}}

While the conservative algorithm strictly avoids the violation of the local 
causality constraint, the optimistic algorithm may process the events that 
are out of the time-stamp order which might violate this constraint. At times 
when the conservative algorithm forces the processors to idle the optimistic 
algorithm enforces the computations and state-updates, thus, according to our 
adopted definition, the theoretical utilization of the optimistic update scheme 
is always at its maximum value of one because the processors never idle. 
However, some of the events in the thus  processed stream of events on a 
processor must be processed prematurely, judging by the random nature of the 
optimistic scheme that takes a risk of guessing whenever the next event is not 
certain. When in the course of an update cycle a processor receives a 
{\it straggler message}, i.e., a message ``from the past" that has its time-stamp 
smaller than the clock value, all the later events that have been processed 
incorrectly must be cancelled. The processor must then send out cancellation 
messages (called {\it anti-messages}) to other processors to {\it roll back} 
all the events that have been processed prematurely. Thus, in the optimistic 
update scheme, although the processors never idle, the computation time of the 
update cycle is not utilized fully efficiently 
because some part of this time is used for the meaningless operations 
(i.e., creation and processing of the rollbacks) and only part of a cycle 
represents the computations that assure the progress of PDES. 

There are many variations of optimistic update schemes, e.g., 
Refs.~ \cite{Jef85,DR90,PS92,Ste93,FC94} and references in \cite{Fuj00}, 
oriented to building implementations with better efficiencies and memory management. 
The key feature of the update mechanism, as described above, and main concepts 
such as rollback and GVT, first introduced in Jefferson's Time Warp \cite{Jef85}, 
can be treated as the generic properties of the optimistic algorithm. In its 
generic form, the algorithm keeps the already processed events in the memory 
in case of the necessary re-processing required by the rollbacks. 

\begin{figure}[tp]
\unitlength1cm
\includegraphics[width=10.0cm]{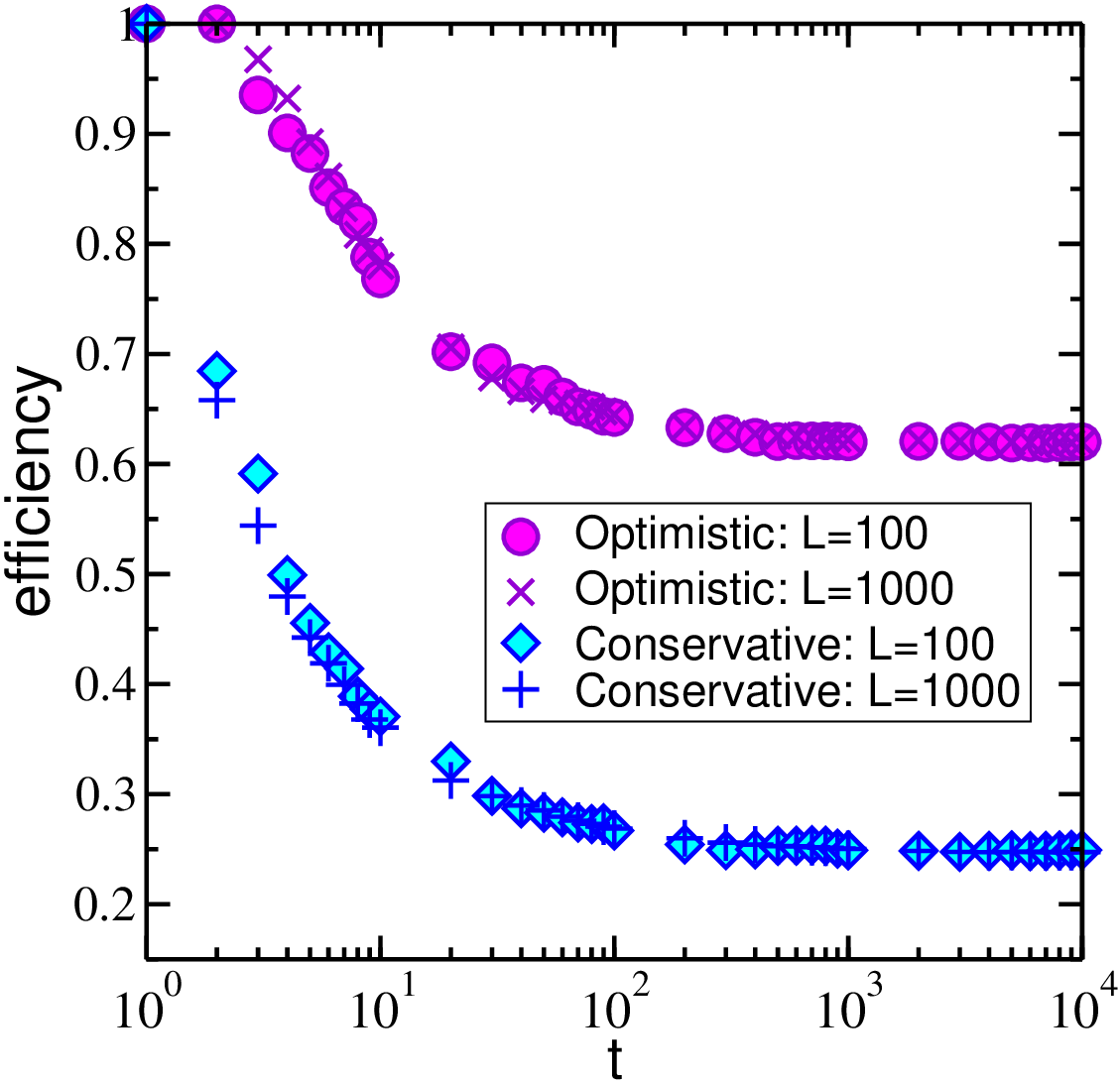}
\caption{\label{kol-11} \baselineskip=0.42truecm
Simulated time-evolution of overall efficiencies in an optimistic 
(upper curve) and a conservative (lower curve) PDES when processors 
carry minimal loads ($L=10,000$).}
\end{figure}

For the ring communication topology, we simulate the growth of the VTH corresponding 
to the optimistic update procedure as Poison-random depositions to the lattice sites 
in analogy to the model described in Sec.~\ref{model}. However, in the optimistic model 
the deposition rule is modified to mimic key features of the optimistic algorithm. 
We assume that each update cycle on each processor consists of processing $N_c$ events, 
only some of which can be eventually committed. With each of these events there is 
the associated random time increment $\eta$. Now the integer index $t$ represents 
the update cycle. The main difference between the optimistic and the conservative 
simulation models is that any time the conservative would wait the optimistic is 
allowed to perform a random guess. In the simplest case of totally unbiased guesses, 
in each cycle the number of correctly guessed events is obtained from a uniform distribution. 
The cumulative simulated LVTs that correspond to processing all events form the 
simulated {\it optimistic} VTH. The cumulative simulated LVTs that correspond to 
processing only the correctly guessed events form the simulated {\it progress} VTH, 
which is embedded in the optimistic VTH. The difference between the optimistic VTH 
and the progress VTH represents the cumulative time that has been wasted by generating 
and processing erroneously guessed events and their associated rollback operations. 
We define the overall efficiency of the optimistic algorithm as the ratio of 
{\it the total progress time} to {\it the  total computation time}. At $t$, the 
total progress time and the total computation time are obtained by integrating 
the progress VTH and the optimistic VTH, respectively, and are represented by 
the areas under these VTH interfaces. In analogy with the above definition, we define 
the overall efficiency of the conservative algorithm as the ratio of the total 
computation time (i.e., the area under the conservative VTH) to the total 
time that the processors spend on computations and idling. These efficiencies are 
presented in Fig.~\ref{kol-11} for the worst-case scenario of the minimal load per 
processor and $N_c=1$. 

Our simulations (Fig.~\ref{kol-11}) confirm the common conception that, 
in the ideal setting, the optimistic algorithm should outperform the conservative 
algorithm \cite{Fuj00}. As Fig.~\ref{kol-11} shows, the lower bound for the 
steady-state conservative efficiency is about $25\%$ and coincides with the 
lower bound obtained for the utilization (Fig.~\ref{kol-03}). For the same case, 
the steady-state optimistic efficiency is about $62\%$; accordingly, the optimistic 
algorithm has a better utilization of the parallelism. In actual applications the 
conservative efficiency can be improved by exploiting in programming a concept of 
lookahead, based on actual properties of the distributed PDES physical model 
under consideration. 

\begin{figure}[tp!]
\includegraphics[width=10.0cm]{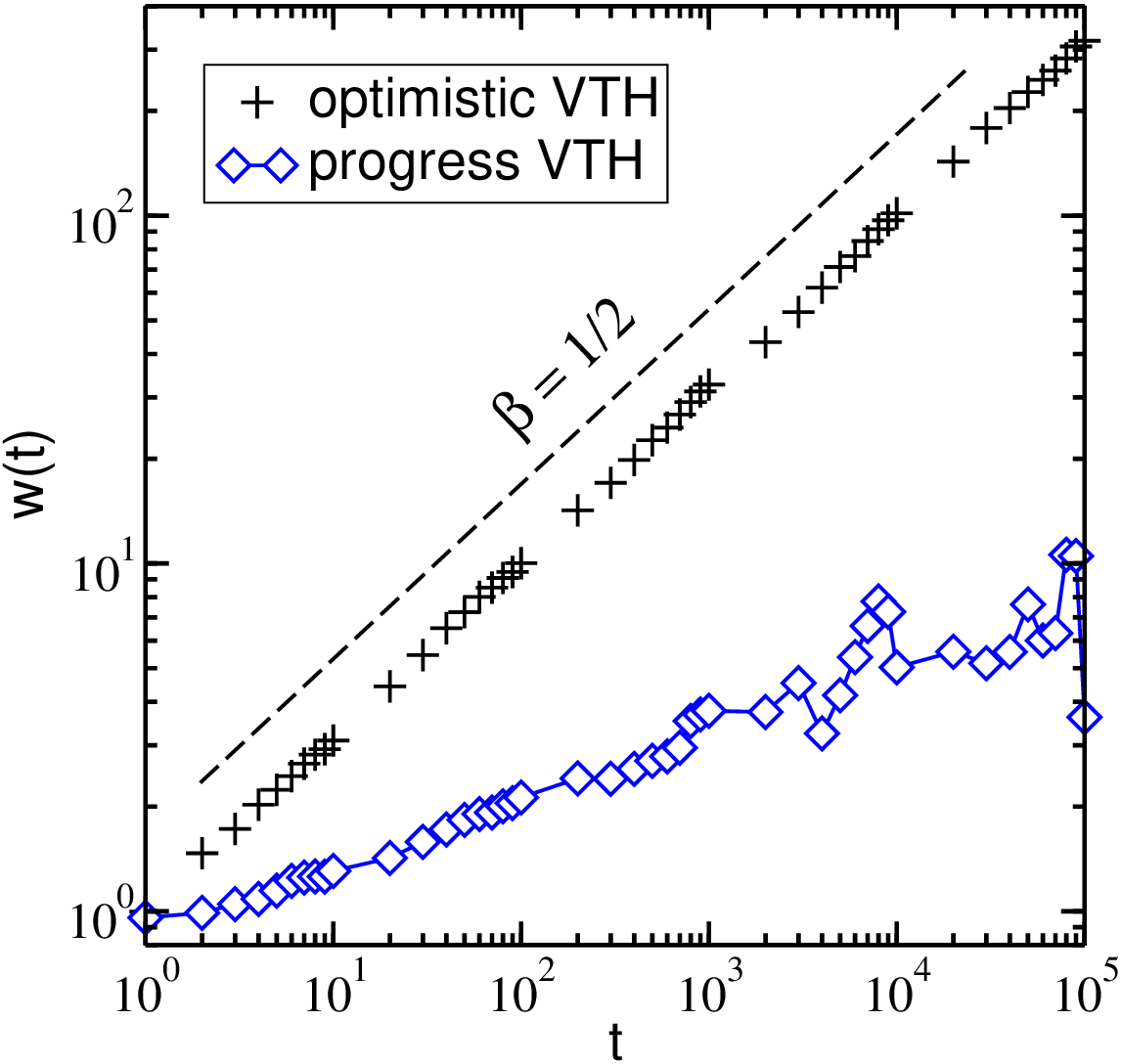}
\caption{\label{kol-12} \baselineskip=0.42truecm
Simulated time-evolution of desynchronizations in an optimistic PDES 
when processors carry minimal loads ($L=10,000$): the widths of the optimistic VTH 
(upper curve, ``plus" symbol) and of the progress VTH (lower curve, diamonds).}
\end{figure}

The statistical spread of the simulated optimistic VTH is presented in 
Fig.~\ref{kol-12} (note, this figure presents the results obtained in only 
one simulation). The simulated optimistic VTH belongs to the RD universality 
class and the spread in local virtual times grows without limit in accordance to 
the power law $w(t) \sim \sqrt{t}$. Intuitively, this result should be expected 
because, by analyzing the operation mode of optimistic updates, one notices that 
the processors work totally independently, progressing their LVTs in an uncorrelated 
fashion. Thus, it should be expected that the memory request per processor required 
to execute generic optimistic PDES grows without bounds as $\sqrt{t}$ when the 
simulations are progressing in $t$. 

The unboundedness of the memory request in a generic optimistic scheme can be also 
justified using quite different arguments \cite{Fuj00}, however, the 
power-law growth for this request, illustrated in Fig.~\ref{kol-12}, has 
been never reported before. The adverse ways in which such an unbounded 
desynchronization affects the performance and standard remedies that can be 
taken to improve on the use of computing resources by optimizing the optimistic 
memory management, are well-known issues (a comprehensive discussion can be 
found in Ref.~\cite{Fuj00}). In general, the generic optimistic update scheme 
requires some kind of explicit synchronization procedure that would limit 
the lengths of rollbacks.

The actual performance of the PDES application may depend on the particulars of 
the underlying physical dynamics of the physical system being simulated and 
the best choice of the algorithm may be uncertain in advance without some 
heuristic trial studies. For example, recently, Overeiner {\it et al.} \cite{OSS02,SOS01} 
reported the first observation of self-organized critical behavior in the performance 
of the optimistic Time Warp algorithm when applied to dynamic Monte Carlo simulations 
for Ising spins on a regular two-dimensional lattice. They found that when this PDES 
approaches a point when the physical Ising-spin phase transition is being simulated 
(i.e., the critical point of the physical dynamics), the average rollback length increases 
dramatically and simulation runtimes increase nonlinearly. 
In Ising-spins simulations, increases in rollback lengths 
are to be expected since around the Ising critical temperature the physical system 
is characterized by the presence of long-range spin-spin correlations and collective 
behavior, where large-scale spin-domains may be overturned simultaneously. 
Consequently, approaching the critical point of physical dynamics should produce 
a decreased number of committed events. However, the simultaneous nonlinear increase 
of the simulation runtime when PDES approaches this critical point seems to be a 
property of the optimistic algorithm since a similar problem is not observed when 
the same physical system is being simulated using the conservative algorithm \cite{KNR99}. 
One possible explanation for this nonlinear deterioration in runtime may be a nonlinear 
cache behavior when rollback lengths increase beyond a certain critical value and 
memory requests increase. The role of the cache behavior, in particular, the nonlinear 
performance degradation with the number of cache misses, has been recently discussed 
in regard with the efficient implementation of an asynchronous conservative protocol 
for a different physical system \cite{GT00}. Another possible explanation, as 
conjectured in Ref.~\cite{OSS02}, may be the onset of self-organized criticality in 
the Time Warp simulation systems, unrelated to the physical critical state. 
Further studies are required to extract universal properties of optimistic protocols 
to identify a class of simulation problems that would show in computations a similar 
behavior to that encountered in PDES for Ising spins.

\section{Conclusion \label{conclude}}

The performance of the distributed PDES algorithms depends in general on  
three main factors: the partitioning of the simulated system among processors, 
the communication topology among the processors, and the update protocol being adopted. 
In a heuristic approach to performance studies (e.g., as in Ref.~\cite{KNR99,OSS02}) 
the application algorithm often utilizes physical properties of the model to be 
simulated; thus, the conclusions of such studies, as being application specific, 
may have a limited scope of generalization. In this chapter, we presented a new 
way to study the performance of PDES algorithms, which makes no explicit reference 
to any particular application. In this new approach, the system of processors 
that perform concurrent update operations in a chosen communication topology is 
seen as a complex  system of statistical physics. First, based on the update and 
the communication patterns of the algorithm, we construct a simulation model for 
its representative virtual time horizon. The statistical properties of this virtual 
time interface correspond to the properties of the algorithm, that is, to the properties 
of the pattern in which the correlations are formed and propagate in the computing system. 
Second, we extract the properties of the algorithm from the statistical properties 
of its simulated virtual time horizon.

In this chapter, we demonstrated how this approach can be used to elucidate the key 
generic properties of the asynchronous conservative parallel update protocol in the 
ring communication topology among processors. For a finite PDES size, i.e., 
a finite load $N$ per processor and a finite number $L$ of processors, 
our findings can be summarized as follows. Both the utilization of the parallel 
processing environment and its desynchronization can be derived explicitly as 
theoretical functions of $L$ and $N$ (these are Eqs.(\ref{util-1}-\ref{util-3}) and 
Eqs.(\ref{evol-2}-\ref{evol-4}), respectively). These functions express 
the existence of the underlying scaling laws for the corresponding virtual time 
horizon and are understood as approximate relations in the sense of statistical 
averages. The existence of these scaling laws presents one aspect of scalability 
of this type of PDES algorithm. The other aspect of algorithmic scalability is 
the behavior of these functions when $N$ and $L$ increase. In the limit of large $L$ 
there is a theoretical non-zero lower bound for the utilization, for any $N$, 
and the value of this bound increases with $N$. On the other hand, for any $N$ and 
$L$ there is a finite upper bound for the desynchronization, thus, for the mean 
memory request per processor during steady-state simulations. Therefore, this kind of 
conservative PDES algorithm is generally scalable.

The model simulation of the virtual time horizon for the generic optimistic update 
protocol in the ring communication topology (Sec.~\ref{compare}) showed that 
the optimistic algorithm lacks the algorithmic scalability. As the optimistic 
simulations evolve in the steady state, the width of the optimistic virtual 
time horizon grows without limit for any finite PDES system. In other words, 
even for the minimal load per processor, the memory request per processor ever 
increases as the square root of the performed number of time-stamped update cycles, 
as the simulations evolve in time. Therefore, the generic version of this algorithm 
demands some form of explicit periodic synchronization.

One advantage of studying the PDES algorithms in terms of their corresponding 
virtual time interfaces is the possibility of deriving explicit diagnostic 
formulas for the performance evaluation, such as, e.g., the evaluation of the 
speedup given by Eqs.(\ref{speedup-2}-\ref{speedup-3}) 
or the estimate of the memory request given by Eq.(\ref{evol-4}) 
for the conservative algorithm considered in this study. These theoretical 
estimates should be treated as the ideal upper bounds for the performance 
when PDES are implemented on the real computing systems. A real implementation 
will produce a deviation from the theoretical prediction, depending on the 
computing platform and on other components of simulation algorithms. The extend 
to which the performance of the implementation scales down from the ideal performance 
should provide important information about possible bottlenecks of the real 
implementation and should be a guide to improving the efficiencies.

The other benefit that comes from the modeling of virtual-time interfaces is a 
relatively inexpensive design tool for new-generation algorithms, without a prior 
need for heuristic studies. For example, knowing that in the ring communication 
topology the maximal conservative memory request per processor for past state savings 
gets larger as the simulation model gets larger, it is easy to predict the maximum 
model size that would fit the available memory in the system. However, the available 
memory resources vary across implementation platforms, so it may happen that one size 
simulation model may fit on one platform and may be too large for the other, having 
the same number of available processors. The question then is: how to modify the update 
algorithm to allow for the tunable memory request? Obviously, the question concerns 
the control of the extent to which the processors get desynchronized in the course of 
simulations, i.e., the control of the VTH width. One can think about a suitable update 
pattern that would model the virtual-time interface of the desired properties and then 
translate this pattern to a new update procedure of the modified algorithm. This 
approach has been used to design a constrained conservative update algorithm \cite{KNK03}, 
where the desynchronization is controlled by the width of a moving virtual update window 
and the ring communication topology is modified to accommodate multiple connections 
between a processor that carries GVT at given update attempt and other processors. 
In another group of conservative algorithms an implicit autonomous synchronization 
may be achieved by modifying the ring communication pattern to accommodate connections 
with the build-in small-world type of communication network \cite{KNGTR03,TKNG03}. 
In both of these modifications, the additional communication network imposed on 
the original ring communication topology serves the sole purpose of reducing the 
desynchronization. Further studies are required in this matter to identify best  
efficient ways of tuning the desynchronization and the memory request.

In summary, the new approach to performance studies, outlined in this chapter, 
that utilizes  simulation modeling of virtual-time interfaces as a tool in 
algorithm design, opens new interdisciplinary research methodologies in the physics of 
complex systems in application to computer science. Promising avenues where 
this kind of approach to complex systems of computer science should lead to 
useful practical solutions may include the criticality issues in distributed 
PDES algorithms, their scalability, prognostication, the design of efficient 
communication networks as well as the development of new diagnostic tools for 
the evaluation of hardware performance.

\vspace{0.43truecm}
\centerline{\bf Acknowledgments}
\vspace{0.43truecm}
This work is supported by 
the ERC Center for Computational Sciences at MSU.
This research used resources of the National Energy Research
Scientific Computing Center, which is supported by the Office
of Science of the US Department of Energy under contract No.
DE-AC03-76SF00098. Partially supported by NSF grants DMR-0113049 and DMR-0426488.


\begin{thebibliography}{99}

\bibitem{Fuj00} Fujimoto R.M., {\it Parallel and Distributed Simulation Systems} 
(John Wiley and Sons, Inc., New York, 2000).

\bibitem{CM79} Chandy K.M. and Misra J., {\it Distributed Simulation: A case 
Study in Design and Verification of Distributed Programs}, IEEE Transactions 
on Software Engineering, 5 (1979), pp.440--452.

\bibitem{Fuj90} Fujimoto R., {\it Parallel Discrete Event Simulation}, 
Communications of the  ACM, 33 (1990), pp.30--53.

\bibitem{Mis86} Misra J., {\it Distributed Discrete-Event Simulation}, 
ACM Computing Surveys,  18 (1986), 39--65.

\bibitem{Lub87} Lubachevsky B.D., {\it Efficient Parallel Simulations of 
Asynchronous Cellular Arrays}, Complex Systems, 1 (1987), pp.1099--1123.

\bibitem{Lub88} Lubachevsky B., {\it Efficient parallel simulation of dynamic 
Ising spin systems}, Journal of Computational Physics, 75 (1988), pp. 103--122.

\bibitem{Jef85} Jefferson D.A., {\it Virtual time}, ACM Transactions on Programming 
Languages and Systems, 7 (1985), pp.404--425.

\bibitem{DR90} Dickens P.M. and Reynolds P.F., {\it SRADS with local rollback}, 
in {\it Proceedings of the SCS Multiconference on Distributed Simulation, San Diego}, 
edited by Nicol D. and Fujimoto R., Simulation Series, 22 (1990), pp.161--164.

\bibitem{PS92} Prakash A. and Subramanian R., {\it An efficient optimistic 
distributed scheme based on conditional knowledge}, in {\it Proceedings of the 
Sixth Parallel and Distributed Simulation Workshop, 1992 SCS Western Multiconference} 
(IEEE Press, New York, 1992), pp.85--96.

\bibitem{Ste93}Steinman J.S., {\it Breathing Time Warp}, in {\it Proceedings 
of the Seventh workshop on Parallel and Distributed Simulation}, edited by 
Bagrodia R. and Jefferson D. (IEEE Computer Society Press, Los Alamitos, CA, 1993), pp.109--118.

\bibitem{FC94} Ferscha, A. and Chiola G., {\it Self Adaptive Logical Processes: 
The Probabilistic Distributed Simulation Protocol}, in {Proceedings of the 
27th Annual Simulation Symposium, LaJolla, 1994} (IEEE Computer Society Press, 
Los Alamitos, CA, 1994), pp.78--88.

\bibitem{KTNR00} Korniss G., Toroczkai Z., Novotny M.A., and Rikvold P.A., 
{\it From Massively Parallel Algorithms and Fluctuating Time Horizons to 
Non-equilibrium Surface Growth}, Physical Review Letters, 84 (2000), pp.1351--1354.

\bibitem{KNTR01} Korniss G., Novotny M.A., Toroczkai Z., and Rikvold P.A., 
{\it Non-equilibrium surface growth and scalability of parallel algorithms 
for large asynchronous systems}, in {\it Computer Simulation Studies in 
Condensed Matter Physics XIII} ed. by Landau D.P., Lewis S.P., and Schuettler H.-B., 
Springer Proceedings in Physics, 86 (Springer-Verlag, 2001), pp.183--188.

\bibitem{KNRGT01} Korniss G., Novotny M.A., Rikvold P.A., Guclu H., and Toroczkai Z., 
{\it Going through rough times: from non-equilibrium surface growth to algorithmic 
scalability}, Materials Research Society Symposium Proceedings Series, 700 (2001), pp.297--308.

\bibitem{KNKG02} Korniss G., Novotny M.A., Kolakowska A., and Guclu H., 
{\it Statistical properties of the simulated time horizon in conservative 
parallel discrete-event simulations}, in {\it Proceedings of the 2002 ACM Symposium 
on Applied Computing, SAC 2002}, (2002), pp.132--138.

\bibitem{KNK03} Kolakowska A., Novotny M.A., and Korniss G., {\it Algorithmic 
scalability in globally constrained conservative parallel discrete-event simulations 
of asynchronous systems}, Physical Review E, 67 (2003), article No 046703, 13 pages.

\bibitem{KNR03} Kolakowska A., Novotny M.A., and Rikvold P.A., {\it Update statistics 
in conservative parallel-discrete-event simulations of asynchronous systems}, 
Physical Review A, 68 (2003), article No 046705, 14 pages.

\bibitem{KNGTR03} Korniss G., Novotny M.A., Guclu H., and Toroczkai Z., 
{\it Suppressing Roughness of Virtual Times in Parallel Discrete-Event Simulations}, 
Science, 299 (2003), pp.677--679.

\bibitem{TKNG03} Toroczkai Z., Korniss G., Novotny M.A., and Guclu H., 
{\it Virtual Time Horizon Control via Communication Network Design}, 
in {\it Computational Complexity and Statistical Physics} ed. by Percus A., 
Istrate G., and Moore C., Santa Fe Institute Studies in the Sciences of 
Complexity Series (Oxford University Press, 2003), in press, ArXiv: cond-mat/0304617.

\bibitem{GKTN04} Guclu H., Korniss G., Toroczkai Z., and Novotny M.A., 
{\it Small-World Synchronized Computing Networks for Scalable Parallel 
Discrete-Event Simulations}, in {\it Complex Networks} ed. by Ben-Naim E., 
Frauenfelder H., and Toroczkai Z., {\it Lecture Notes in Physics} (Springer, 2004), in press.

\bibitem{LB00} Landau D.P. and Binder K., {\it A Guide to Monte Carlo Simulations in 
Statistical Physics} (Cambridge University Press, Cambridge, 1995).

\bibitem{KNR99} Korniss G., Toroczkai Z., Novotny M.A., and Rikvold P.A., 
{\it Parallelization of a Dynamic Monte Carlo Algorithm: a Partially Rejection-Free 
Conservative Approach}, Journal of Computational Physics, 153 (1999), pp.488--508.

\bibitem{NKK03} Novotny M.A., Kolakowska A., and Korniss G., {\it Algorithms for Faster 
and Larger Dynamic Metropolis Simulations}, in {\it The Monte Carlo Method in the Physical 
Sciences}, ed. by Gubernatis J.E, AIP Conference Proceedings, Vol. 690 (American 
Institute of Physics, New York, 2003), pp. 240--247.

\bibitem{BS95} Barabasi A.-L. and Stanley H.E., {\it Fractal Concepts in Surface Growth}
(Cambridge University Press, Cambridge, 1995).

\bibitem{FV85} Family F. and Vicsek T., {\it Scaling of the active zone in the Eden 
process on percolation networks and the ballistic deposition model}, 
Journal of Physics A, 18, (1985), pp.L75--L81.

\bibitem{KPZ86} Kardar M., Parisi G., and Zhang Y.-C., {\it Dynamic scaling 
of growing interfaces}, Physical Review Letters, 56 (1986), pp.889--892.

\bibitem{EW82} Edwards S.F. and Wilkinson D.R., {\it The Surface Statistics of 
a Granular Aggregate}, Proceedings of the Royal Society London A, 381 (1982), pp.17--31.

\bibitem{KN04} Kolakowska A. and Novotny M.A., {\it Discrete-event analytic technique 
for surface growth problems}, Physical Review B, 69 (2004), article No 075407, 5 pages.

\bibitem{KNV04} Kolakowska A., Novotny M.A., and Verma P.S., 
{\it Roughening of the interfaces in $(1+1)$ dimansional two-component surface growth with an 
admixture of random deposition}, Physical Review E, in press (2004), 16 pages, 
ArXiv: cond-mat/0403341.

\bibitem{OSS02} Overeiner B. J., Schoneveld A., and Sloot P. M. A., 
{\it Self-Organized Criticality in Optimistic Simulations of Correlated Systems}, 
in {\it Parallel and Distributed Discrete Event Simulation}, edited by Tropper C. 
(Nova Science Publishers, New York, 2002), pp.79-98.

\bibitem{SOS01} Sloot P. M. A., Overeiner B. J., and Schoneveld A., 
{\it Self-Organized Criticality in Simulated Correlated Systems}, 
Computer Physics Communications, 142 (2001), pp.76-81.

\bibitem{GT00} Gan B.-P. and Turner, S. J., {\it An Asynchronous Protocol 
for Virtual Factory Simulation on Shared Memory Multiprocessor Systems}, 
Journal of Operational Research Society, Special Issue on Progress in Simulation 
Research, Vol. 51, No. 4 (2000), pp.413-422.

\end{thebibliography}
\end{document}